\documentclass[cernpreprint,atlasdraft=false,texlive=2016,american,PAPER,texmf]{atlasdoc}
 
\AtlasJournal{Phys.\ Rev.\ D}
 
\usepackage{atlaspackage}
\usepackage{atlasbiblatex}
 
\usepackage{atlascontribute}
 
\usepackage{atlasphysics}
 
\graphicspath{{logos/}{newplots/}}

\usepackage{float}

\AtlasTitle{Measurement of the relative $B^{\pm}_{c}/B^{\pm}$ production cross section with the ATLAS detector at $\sqrt{s}=8$ TeV}

\AtlasAbstract{
The total cross section and differential cross sections for the production of
$B_{c}^{\pm}$ mesons, times their branching fraction to $J/\psi
\pi^{\pm}$,  are measured relative to those for the production of
$B^{\pm}$ mesons, times their
branching fraction to $J/\psi K^{\pm}$. The data used for
this study correspond to an integrated luminosity of
$20.3\,\mathrm{fb}^{-1}$ of $pp$ collisions recorded by the ATLAS
detector at the Large Hadron Collider in 2012 at a center-of-mass
energy of $\sqrt{s} = 8$ TeV.
The measurement is performed differentially in
bins of transverse momentum $p_{\mathrm{T}}$ for
$13$~GeV~$<p_{\mathrm{T}}(B_{c}^{\pm})<22$~GeV and
$p_{\mathrm{T}}(B_{c}^{\pm}) > 22$~GeV and in bins of
rapidity $y$ for $|y| <0.75$ and $0.75<|y|<2.3$.
The relative cross section times branching fraction for
the full range $p_{\mathrm T} > 13$ GeV and $|y| < 2.3$ is
$(0.34\,\pm\,0.04_{\text{stat}}\,^{+0.06}_{-0.02}\,_{\text{sys}}\,\pm\,0.01_{\text{lifetime}})\%$.
The differential measurements suggest that the production cross
section of the $B^{\pm}_c$ decreases faster with $p_{\mathrm T}$
than the production cross section of the $B^{\pm}$, while no
significant dependence on rapidity is observed.
}
 
\author{The ATLAS Collaboration}

\AtlasRefCode{BPHY-2016-03}
 
\AtlasNote{BPHY-2016-03}
 
\PreprintIdNumber{CERN-EP-2019-194}
\hypersetup{pdftitle={ATLAS document},pdfauthor={The ATLAS Collaboration}}
 
\begin{document}
 
\maketitle

\section{Introduction}
\label{sec:intro}
 
The $B^{\pm}_c$ meson is a bound state of the two heaviest distinct
quarks able to form a stable state, $c$ and $\bar b$ for $B_c^+$ or
$\bar c$ and $b$ for $B_c^-$.
Measurements of its production can provide unique insight into
heavy-quark hadronization: unlike lighter $B$-states, production of
a $B^{\pm}_c$ meson requires collinear production of two distinct
heavy quarks.
 
A measurement of the production cross section times
branching fraction for $B_{c}^{\pm} \to J/\psi\pi^{\pm}$,
relative to that for $B^{\pm} \to J/\psi K^{\pm}$, at
$\sqrt{s}=8\,\TeV$ of $pp$ collisions and differential in transverse momentum
$p_{\mathrm T}$ and in rapidity\footnote{The rapidity $y$ is defined as
$(1/2)\times \ln[ ( E + p_z ) / ( E - p_z ) ]$, where
$E$ is the energy of a particle and $p_z$ is its component
of momentum along the beam axis.} $y$,
has not yet been reported for the fiducial region considered in this
article and defined below, although measurements of the
individual cross sections have been published.
At $\sqrt{s}=7\,\TeV$, the LHCb Collaboration reported \cite{LHCb} a
relative cross section times branching fraction of
$(0.68\,\pm\,0.10_{\mathrm{stat}}\,\pm\,0.03_{\mathrm{syst}}\,\pm\,0.05_{\mathrm{lifetime}})\%$
for $p_{\mathrm{T}}(B_{c}^{\pm})>4\,\GeV$ in the
pseudorapidity\footnote{
The pseudorapidity $\eta$ is defined as
$-\ln[\tan(\theta/2)]$, where $\theta$ is the angle between the
particle momentum and the positive direction of the beam axis.
}
range $2.5<\eta<4.5$. LHCb reported \cite{LHCb2} for the same
differential relative $B^{+}_c$ cross-section at $\sqrt{s}=8$\,\TeV\
a measurement of $(0.683 \pm 0.018_{\text{stat}} \pm 0.009_{\text{syst}})\%$
in the rapidity range $2.0 < y < 4.5$ for the interval
$0 < p_{\mathrm T}(B_c^{+}) < 20$\,\GeV. The measurement \cite{LHCb2} by the
LHCb Collaboration at $\sqrt{s}=8$\,\TeV\ has a very slight overlap
with the fiducial region (defined below) of this one. In the fiducial
region $13 < p_{\mathrm T}(B) < 20$\,\GeV\ and $2.0<y(B)<2.3$, where the
two experiments are both sensitive, the LHCb data show no trend in the
ratio as a function of $p_{\mathrm T}$. However, the LHCb dataset is
dominated by $B_c$ candidates with $p_{\mathrm T}$ lower than those in
this ATLAS analysis. The CMS Collaboration measured the same quantity in the rapidity range
$|y|<1.6$ for $p_{\mathrm{T}}(B^{\pm}_c)>15\,\GeV$ to be
$(0.48\,\pm\,0.05_{\mathrm{stat}}\,\pm\,0.03_{\mathrm{syst}}\,\pm\,0.05_{\mathrm{lifetime}})\%$
\cite{CMS} at $\sqrt{s}=7$\,\TeV.
The values reported here by ATLAS are smaller than
those obtained by the other experiments at different energies and with
different fiducial volumes, and evidence of dependence of this
relative production cross section on transverse momentum is shown here
for the first time.

Theoretical predictions for the hadronic $B_{c}^{\pm}$ production
cross section have been reported by several authors \cite{Masetti,
Berezhnoy, Chang, Gouz, Slabospitsky, Berezhnoy2, Chang2} and this
study is motivated by the range of predicted
values. Theoretical calculations predict that the total production
cross section $\sigma_{B_{c}}$ is in the range of $1\,\mathrm{nb}$ to
$31.5\,\mathrm{nb}$ at $\sqrt{s}=1.8\,\TeV$ and
$44\,\mathrm{nb}$ to $190\,\mathrm{nb}$ at
$\sqrt{s}=14\,\TeV$, but no published calculation for the
relative cross section at $\sqrt{s}=8\,\TeV$ is available at
this time. These calculations depend on the square of the decay constant
$f_{B_{c}^{\pm}}$ \cite{Masetti, DC1, DC2}, which in the
non-relativistic approach is proportional to the absolute value of the
wave function at the origin.
 
In this measurement of the cross section times branching fraction for
$B_{c}^{\pm} \to J/\psi\pi^{\pm}$
relative to that for $B^{\pm} \to J/\psi K^{\pm}$, the $J/\psi$ mesons
are reconstructed from their decays into $\mu^{+}\mu^{-}$. The relative
cross-section measurement is differential in two
bins of the transverse momentum $p_{\mathrm T}$ of the $B^{\pm}_c$,
$13\,\GeV < p_{\mathrm T}(B^{\pm}_c) < 22\,\GeV$ and $p_{\mathrm
T}(B^{\pm}) > 22\,\GeV$, for rapidity $|y| < 2.3$, and in two bins
of rapidity y of the $B^{\pm}_c$, $|y| < 0.75$ and $0.75 < |y| < 2.3$,
for $p_{\mathrm T}(B^{\pm}_c) >13\,\GeV$.
 
The relative cross section is also measured for the inclusive dataset with
$p_{\mathrm T} > 13$\,\GeV\ and $|y| < 2.3$.
The fiducial volume is defined by the requirements
on  $p_{\mathrm{T}}(B^{\pm}_c)$ and $y(B^{\pm}_c)$. The bins are
selected to equalize yields of the $B^{\pm}_c$. The same bin sizes are
used for the $B^{\pm}_c$ and the $B^{\pm}$.
The data were recorded in 2012 by the ATLAS \cite{ATLAS_2008}
experiment at a center-of-mass energy of $\sqrt{s}=8\,\TeV$.

The article is organized as follows. After this introduction,
Section~2 provides an overview of the ATLAS detector including the
trigger system and the objects used. Section~3
describes the simulation, reconstruction, and event
selection. Section~4 shows the relative cross-section
calculation. Section~5 presents the fits to the mass
distributions. Section~6 describes the calculation of the efficiency and acceptance
ratios. Section~7 reports the systematic uncertainties. Section~8
presents the results of the measurements. Conclusions are drawn
in Section~9.
 
\section{The ATLAS detector, trigger, and basic track selections}
\label{sec:detector}
 
ATLAS is a general-purpose particle detector
\cite{ATLAS_2008} consisting of several subsystems including the inner
detector (ID), the electromagnetic and hadronic calorimeters, and the
muon spectrometer (MS). Muons pass through the calorimeters and reach
the MS if their momentum is above approximately $3\,\GeV$. The
ID features a three-component tracking system, consisting of two
silicon-based detectors, the pixel detector and the microstrip
semiconductor tracker (SCT), and the transition radiation tracker
(TRT).
 
ID tracks are reconstructed if their transverse
momentum is greater than
$400\,\MeV$ and if the magnitude of their pseudorapidity,
$|\eta|$, is less than $2.5$. Muon candidates are either formed from a
stand-alone MS track that is matched to an ID track (so-called
combined muons) or from an ID track extrapolated to the MS and matched
to track segments in the MS (so-called tagged muons)
\cite{MuonReco}. The ID and MS subsystems are of particular importance
to this study. Only the data taken when both of these subsystems were
properly operational and when LHC beams were stable are used,
corresponding to an integrated luminosity of $20.3\,\mathrm{fb}^{-1}$.
Although
both the ID and the MS provide momentum measurements, in the
$p_{\mathrm T}$ range relevant to this analysis, the MS momentum
resolution is worse than that of the ID due to energy loss in the
calorimeters. Therefore, the MS is used only to identify muons, and the
$p_{\mathrm T}$ measurement is taken from the ID. The muon candidates
are required to have the number of pixel hits plus the number of
crossed dead pixel sensors greater than zero, the number of SCT hits plus
the number of crossed dead SCT sensors greater than four, and the number
of pixel holes\footnote{In this context, a hole is counted when a hit
is expected in an active sensor located on the track trajectory
between the first and the last hit associated with this track, but
no such hit is found. If the corresponding sensor is known to be
inactive and therefore not expected to provide a hit, no hole is
counted.} plus the number of SCT holes on the track less than three.

ID tracks from charged hadrons are required to have at least two pixel hits and at
least six hits in the SCT (eight hits in total). If a track crosses a dead
sensor, it is not counted as a hit in the corresponding subdetector;
the number of pixel+SCT holes is required to be less than
three. For tracks with $0.1 < |\eta| < 1.9$ it is required that $n > 5$
and $n^{\text{out}}_{\text{TRT}} < 0.9n$, where $n = n_{\text{TRT}} +
n^{\text{out}}_{\text{TRT}}$, and
$n_{\text{TRT}}\, (n^{\text{out}}_{\text{TRT}})$ stands for the number
of TRT hits (outliers \cite{MuonReco}) on the track.
 
The trigger system \cite{Trigger} for data collected up to and
including 2012 comprises three levels: the hardware-based first-level (L1)
trigger and the high-level triggers (HLT), consisting of the second-level
trigger and the event filter. The L1 trigger uses resistive
plate chambers and thin gap chambers to trigger on muons in the
pseudorapidity ranges of $|\eta|<1.05$ and $1.05<|\eta|<2.5$,
respectively. One or more regions-of-interest (RoI), identified by the
L1 muon trigger, seed the HLT muon reconstruction
algorithms, where the tracks from both the ID and the MS are
combined. For this analysis the HLT selection of the $J/\psi$ requires
two oppositely charged muons, originating at
a common vertex, with the invariant mass of the muon pair lying
between $2.5\,\GeV$ and $4.3\,\GeV$. The individual
muon $p_{\mathrm{T}}$ thresholds are both 4\,\GeV. The pseudorapidity
range of the muon selection is $|\eta|< 2.5$.

\section{Simulation, reconstruction, and event selection}
Samples for the signal ($B^{\pm}_c\to J/\psi \pi^{\pm}$ and
$B^{\pm}\to J/\psi K^{\pm}$) were generated with the \textsc{Pythia}
Monte Carlo (MC) generator \cite{Pythia6, Pythia8}. For the generation
of the $B^{\pm}_c$ meson a dedicated \textsc{Pythia} extension was used, based
on theoretical input \cite{Berezhnoy, Berezhnoy2}. The
parameters of the MC simulation are described in Section~6.
The response of the ATLAS detector was simulated with the ATLFAST2
procedure \cite{ATLAS_Sim} using the \textsc{Geant4} package
\cite{Geant4}, and the events are reconstructed with the same
software as is used to process the data from the detector. The effect
of multiple proton--proton collisions per beam crossing, also known as
pileup, is included.
 
The $J/\psi$ candidates are reconstructed from pairs of oppositely charged
muon candidates. To allow the most accurate corrections for trigger
efficiencies, each reconstructed muon candidate is
required to match a trigger-identified muon candidate within a
cone\footnote{ATLAS uses a right-handed coordinate system
with its origin at the nominal interaction point (IP) in the center
of the detector and the $z$-axis along the beam line. The $x$-axis
points from the IP to the center of the LHC ring, and the $y$-axis
points upward. Cylindrical coordinates $(r,\phi)$ are used in the
transverse plane, $\phi$ being the azimuthal angle around the $z$-axis
and $r$ the distance from the IP in the transverse plane.}
of size $\Delta R = \sqrt{(\Delta \eta)^2 + (\Delta \phi)^2} = 0.01$.
The efficiency of matching the reconstructed muon candidates to the
trigger muon candidates is comparable for the two $B$-meson species
according to MC simulation.
The muon pair is fitted to a common vertex with the requirement that at
least one of the muons is a combined muon.  The procedure is described
in Ref.~\cite{Fitting}.
The following requirements are applied to choose a
$J/\psi$ candidate:
\begin{itemize}
\item The $p_{\mathrm T}$ of each of the two muon candidates
must be above 4\,\GeV.
\item The vertex fit quality $\chi^2 / (N_{\text{d.o.f.}} = 1)$ must
be below 15. This soft requirement is chosen to remove highly unlikely
muon combinations.
\item The invariant mass $m(\mu^+\mu^-)$ calculated from the
re-fitted track parameters must
lie in a mass window that depends on the pseudorapidity of the
muons. The $\eta$ ranges are determined from the resolution of the muon
detectors.
Three mass windows are defined for the $J/\psi$ around the world average value
of 3096.916\,\MeV\ \cite{PDG_2018}, as follows.  If the $|\eta|$ of both
muons is less than (greater than) $1.05$, the mass window is $\pm 120$\,\MeV\
($\pm 270\,\MeV$).  If the $|\eta|$ of one muon is
less than 1.05 and the $|\eta|$ of the other muon is between 1.05
and 2.5, the mass window is  $\pm 180$\,\MeV.
 
\end{itemize}
The muon mass is assigned to each track in the $J/\psi$ reconstruction. The
re-fitted track parameters are derived from the vertex fit.
 
The $B$ candidates (where $B$ represents $B_{c}^{\pm}$ or $B^{\pm}$) are
reconstructed by fitting the
tracks of the two muons from the $J/\psi$ candidate, together with a
charged-hadron track, to a common vertex. For the $B^{\pm}$ candidates
the kaon mass is assigned to the hadron track while for the $B^{\pm}_c$ candidates
the pion mass is assigned to the hadron track. The mass of the $J/\psi$
meson is constrained in the fit to its world average value.
 
The primary $pp$ interaction vertex is found by extrapolating the flight direction
of the reconstructed $B$ candidate to the $z$-axis and then selecting the closest
vertex in the $z$ direction \cite{PERF-2015-01}.  The selected vertex is re-fitted with the
three signal tracks (two muonic and one hadronic) removed.
 
A significant part of the combinatorial background consists of
$J/\psi$ mesons that combine with light hadrons that are
not associated with the decay. This
background can be reduced by applying a minimum value for the significance of the impact parameter
of the hadron track relative to the primary vertex in the transverse plane, $d^0_{xy}/\sigma (d^0_{xy})$, where
$d^0_{xy}$ is the projection of the impact parameter onto the
transverse plane and $\sigma(d^0_{xy})$ is the
uncertainty of $d^0_{xy}$.
 
Another significant source of background which is especially important
for the $B_c^{\pm}$ is the partially reconstructed semileptonic decays
of the $B_c^{\pm}$ such as $B^{\pm}_c \to J/\psi \mu^{\pm} \nu_{\mu}$. This
background is reduced by removing combinations in which one of the
hadronic candidates is identified as a muon by the MS.

Reconstructed $B$ candidates are required to
satisfy the following selection criteria:
\begin{itemize}
\item The $\chi^2 / (N_{\text{d.o.f.}} = 4)$ of the fit of the $B$ vertex must be below 1.8.
\item The rapidity $|y(B)|$ must be less than 2.3.
\item The $p_{\mathrm{T}}$ of the hadron
candidate must be above 2.0\,\GeV.
\item To suppress combinatorial background due to prompt\footnote{``Prompt" here
refer to states that are produced directly by short-lived QCD sources, i.e.
those not produced in the beauty hadron decays.} $J/\psi$
candidates, the impact parameter significance $d^{0}_{xy}/\sigma(d^{0}_{xy})$
of the hadron candidate must exceed 1.2. MC studies demonstrated that this
criterion is more efficient for the $B^{\pm}_c$ candidate selection than
a requirement on the lifetime.
 
\end{itemize}
 
In events with multiple $B$ candidates, the candidate with the
best $\chi^2$ from the vertex fits is used. The fraction of
multiple-candidate events is negligible for this analysis, as observed
in data and confirmed by MC studies. This analysis uses all
the ground-state $B$ candidates including those produced directly and
those produced from the cascade decay of excited states.
 
\section{Relative cross-section calculation}
The relative cross section times branching fraction is given by:
\begin{equation}\frac{\sigma(B^{\pm}_{c})\cdot \mathcal{B}(B_{c}^{\pm}\to
J/\psi \pi^{\pm})\cdot \mathcal{B}(J/\psi \to
\mu^{+}\mu^{-})}{\sigma(B^{\pm})\cdot \mathcal{B}(B^{\pm}\to J/\psi
K^{\pm})\cdot \mathcal{B}(J/\psi \to
\mu^{+}\mu^{-})}=\frac{N^{\text{reco}}({B_{c}^{\pm}})}{N^{\text{reco}}({B^{\pm}})}\cdot\frac{\epsilon(B^{\pm})}{\epsilon(B^{\pm}_{c})}.\label{eqn:crossSection}\end{equation}
The notation $N^{\text{reco}}(X)$ refers
to the number of reconstructed collision data events, where $X$ is
either $B_{c}^{\pm}$ or $B^{\pm}$.
The $\epsilon(B^{\pm})$ and $\epsilon(B^{\pm}_{c})$ are the
efficiencies of $B^{\pm}$  and  $B_c^{\pm}$ reconstruction that
correct the numbers $N^{\text{reco}}({B_{c}^{\pm}})$
and $N^{\text{reco}}({B^{\pm}})$  for detector effects, selection
criteria, differences between interactions of $K^{\pm}$ and $\pi^{\pm}$
with the detector material, as well as efficiencies associated with the
trigger.

\section{Fit to the mass distributions}\label{sect:yield}
Extended unbinned maximum-likelihood fits to the mass distributions of the $B_{c}^{\pm}$ and the $B^{\pm}$ are performed to extract $N^{\text{reco}}({B_{c}^{\pm}})$ and $N^{\text{reco}}({B^{\pm}})$ from the data in each bin in $p_{\mathrm{T}}$ and $|y|$.  This involves calculating the parameters that maximize the {likelihood function}, defined as
\begin{equation*} \mathcal{L} = \frac{e^{-N_{\mathrm{sig}}-N_{\mathrm
{bkg}}}}{N!}\prod^N_{i=1} \left [N_{\mathrm {sig}}
\mathcal{F}_{\mathrm{signal}}(m^i_{J/\psi X_h}, \delta m^{i}_{J/\psi X_h}) + N_{\mathrm{bkg}}\mathcal{F}_{\mathrm{bkg}}(m^i_{J/\psi X_h}) \right
] \label{eqn:likelihoodFunc}\end{equation*}
where $N$ is the total number of $J/\psi X_h$ candidates, $X_h$ represents the corresponding hadron,
$N_{\mathrm{sig}}$ is the total number of signal events, and $N_{\mathrm{bkg}}$ is the
total number of background events. The contribution from the signal
$\mathcal{F}_{\mathrm{signal}}$ is modeled by a Gaussian
probability density function with event-by-event errors. It is given for the $B_{c}^{\pm}$ by
\begin{equation*}\mathcal{F}_{\mathrm{signal}}(m^{i}_{J/\psi \pi}, \delta m^{i}_{J/\psi \pi})\propto
\exp\left(-\frac{(m^{i}_{J/\psi \pi^\pm} -m_{B_{c}^{\pm}})^{2}}{2\,(s_1\,\delta m^{i}_{J/\psi \pi^\pm})^2}\right)  \label{eqn:gausDensBc}\end{equation*}
and for the $B^{\pm}$ by
\begin{equation*}\mathcal{F}_{\mathrm{signal}}(m^{i}_{J/\psi K^{\pm}}, \delta m^{i}_{J/\psi K^{\pm}})\propto
\exp\left(-\frac{(m^{i}_{J/\psi K^{\pm}} -m_{B^{\pm}})^{2}}{2\,(s_2\,\delta m^{i}_{J/\psi K^{\pm}})^2}\right)  \label{eqn:gausDensBp}\end{equation*}
where $m_{B^{\pm}}$ and $m_{B_{c}^{\pm}}$ are the masses of the
$B^{\pm}$ and $B_{c}^{\pm}$, respectively. The variables $m_{B^{\pm}}$
and $m_{B_c^{\pm}}$ are taken as free parameters in the fit.
The widths $s_{1}\,\delta m^{i}_{J/\psi \pi^{\pm}}$ and $s_{2}\,\delta
m^{i}_{J/\psi K^{\pm}}$ are the products of the corresponding scale
factors $s_1$ and $s_2$ and
the event-by-event mass resolution. The event-by-event mass uncertainty
is calculated from the tracking covariance matrices by error propagation.
The scale factors are free parameters of the fit which account for
imperfection in estimates of the mass errors.
Ideally the value of $s_{1,2}$ is one.
The mass resolution $\sigma_m$ is obtained from the fits and
appears in the relevant tables below. It is defined as the half-width
of the region of the $J/\psi$ mass distribution for
which the integral of the sum of
$\mathcal{F}_{\mathrm{signal}}(m_{J/\psi X_h},
\delta(m_{J/\psi X_h}))$ over all
candidates contains 68.27\% of $N_{\mathrm{sig}}$.
 
The background to $B_{c}^{\pm}$ production is modeled with an exponential function plus a constant term,
\begin{equation*}\mathcal{F}_{\mathrm{bkg}}\propto \exp(a\cdot m^{i}_{J/\psi \pi^{\pm}})+b .\label{eqn:backgroundExp}\end{equation*}
 
In the $B^{\pm}$ mass region, partially reconstructed $b$-hadron decays
populate the lower part of the $B^{\pm}$ mass spectrum. Their contributions
are estimated with a complementary error function
\begin{equation*} \mathcal{F}^{(1)}_{\mathrm{bkg}}(m^{i}_{J/\psi
K^{\pm}}) \propto 1 - \mathrm{erf}\left(\frac{m^{i}_{J/\psi
K^{\pm}}-m_{0}}{s_{0}}\right) = 1 - \frac{2}{\sqrt{\pi}}\int\limits_{0}^{A}e^{-t^{2}}\,dt\label{eqn:errFunc}\end{equation*}
where $A = \frac{m^{i}_{J/\psi K^{\pm}} -m _{0}}{s_{0}} $, and $m_{0}$ and $s_{0}$ determine the position and the slope of the error function, respectively.
 
The Cabibbo-suppressed decay $B^{\pm}\rightarrow J/\psi \pi^{\pm}$
populates the high part of the $B^{\pm}$ mass spectrum. It is modeled by a Gaussian function
\begin{equation*} \mathcal{F}^{(2)}_{\mathrm{bkg}}(m^{i}_{J/\psi K^{\pm}}) \propto \exp\left(-\frac{(m^{i}_{J/\psi K^{\pm}} -m_{B^{\pm},\pi^{\pm}})^2}{2\,(s_{3}\,\delta m^{i}_{J/\psi K^{\pm}})^2}\right), \label{eqn:multiGauss1}\end{equation*}
where $s_3$ is the corresponding scale factor.
The variable $m_{B^{\pm},\pi^{\pm}}$ is the median of the mass distribution
in the data for  $B^{\pm} \to J/\psi \pi^{\pm}$ events when the kaon mass
instead of the pion mass is assumed for the charged-hadron track.
 
The remaining background is mostly due to production of $J/\psi$
mesons from decays of $b$-hadrons other than the $B^{\pm}$, which are
combined with a random hadron track. They are described with an
exponential function
\begin{equation}\mathcal{F}^{(3)}_{\mathrm{bkg}}(m^{i}_{J/\psi K^{\pm}})
\propto \exp\left(d\cdot m^{i}_{J/\psi
K^{\pm}}\right),\label{eqn:expFunc1}\end{equation}
where $d$ is a constant.
 
The models for the background contributions are combined with
different fractions, which are fitted to the data. The relative fractions
are left free in the fit, as ($a \times \mathcal{F}^{(1)}_{\mathrm{bkg}}
+ b \times \mathcal{F}^{(2)}_{\mathrm{bkg}} +
(1-(a+b)) \times \mathcal{F}^{(3)}_{\mathrm{bkg}}$), where $a$ and $b$
are free parameters of the fit.
 
\begin{figure}[H]
\centering
\begin{tabular}{c}
\includegraphics[width=8.2 cm]{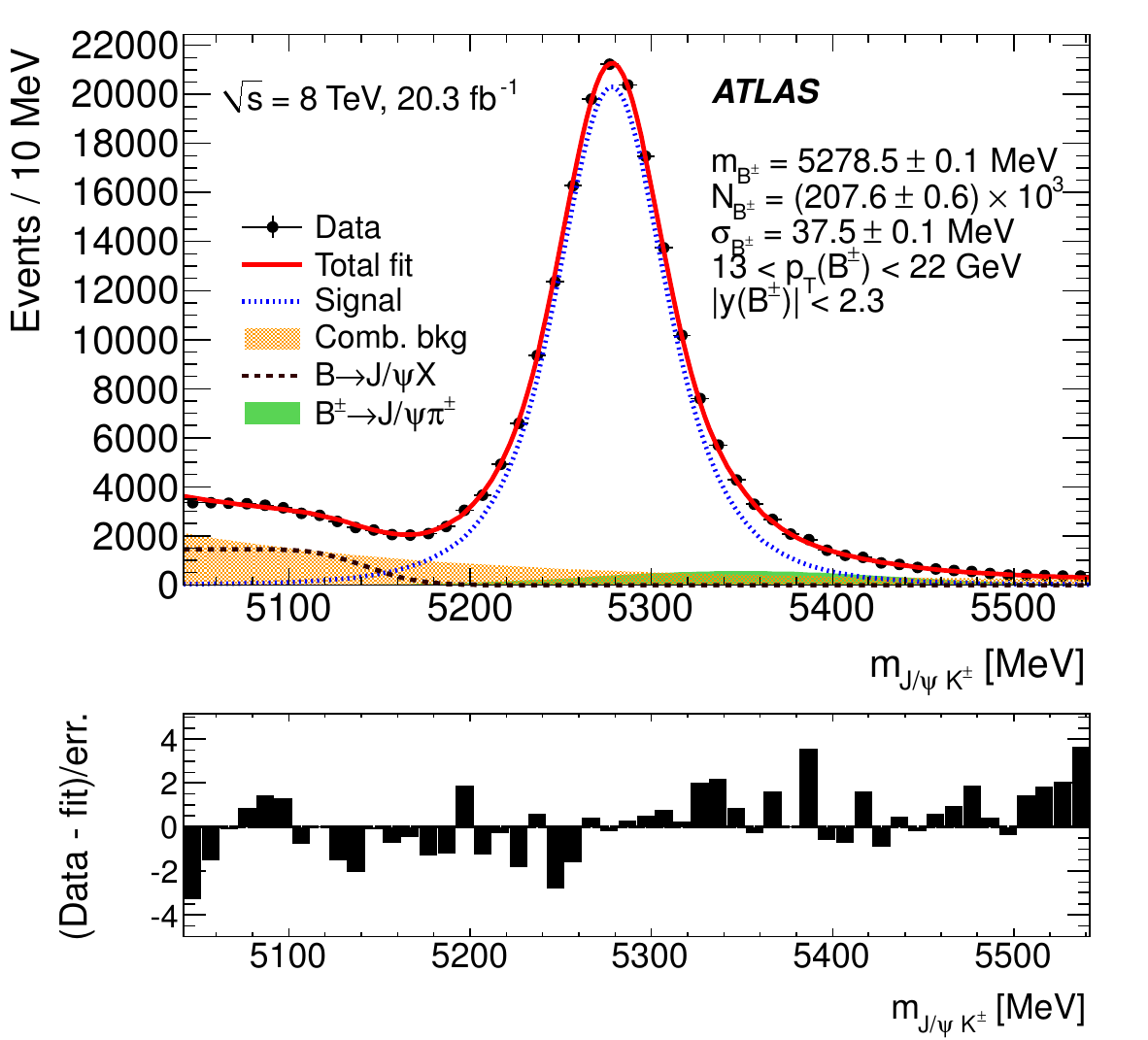} \\
\includegraphics[width=8.2 cm]{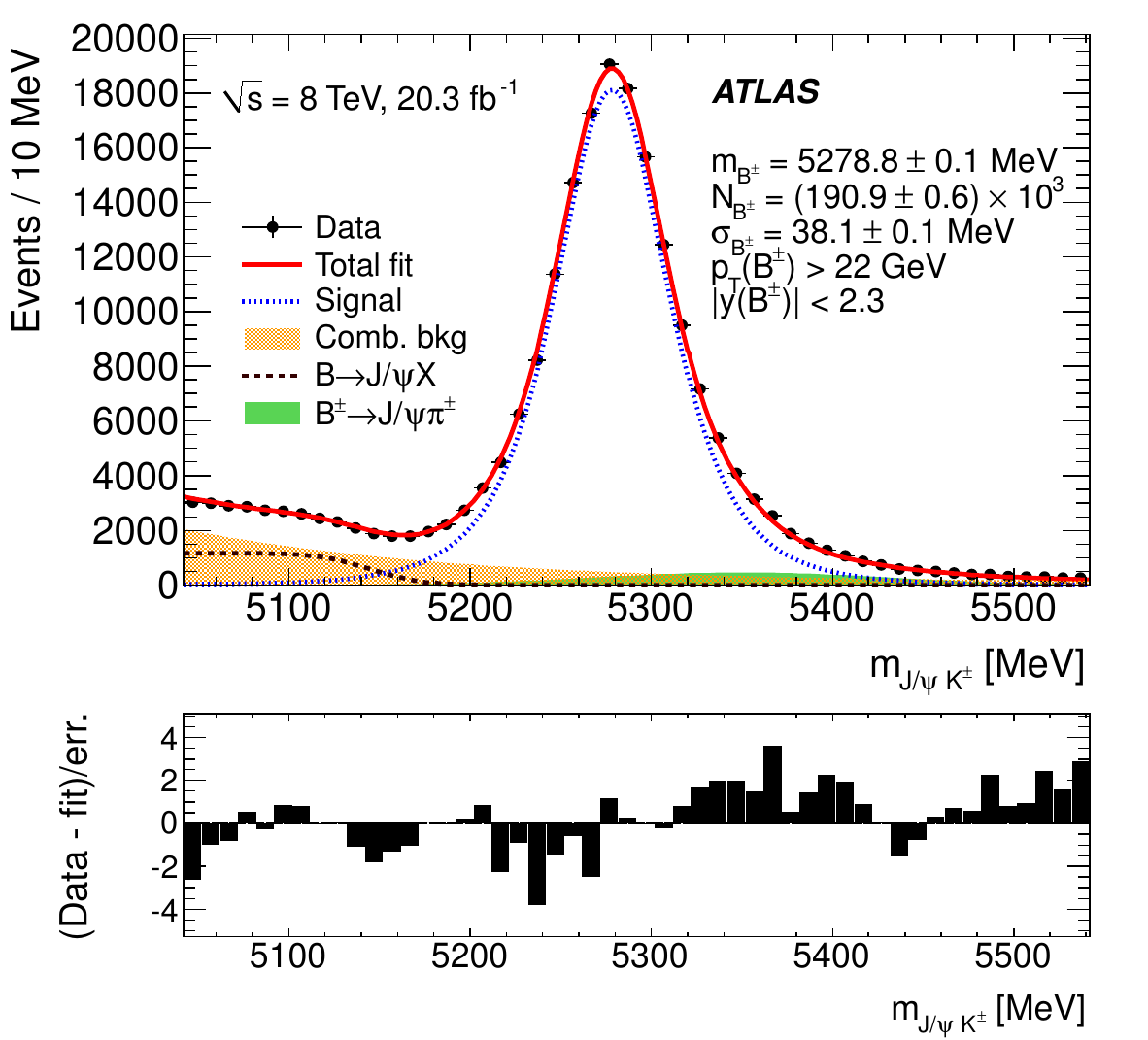} \\
\end{tabular}
\caption[The $B^{\pm}$ mass fit in two $p_{\mathrm{T}}(B^{\pm})$
bins.]{The projection of the fit of the invariant mass distribution for the $B^{\pm}$
meson. ``Signal'' stands for the signal component of the
fit, ``Comb.\ bkg'' stands for the combinatorial background
component of the fit, and ``Total fit'' stands for the fit to
the sum of the signal and all background components. The fit for the low
transverse momentum bin
($13\,\GeV < p_{\mathrm{T}}(B^{\pm}) < 22\,\GeV$) is
shown at the top, and that for the high transverse momentum bin
(${p_{\mathrm{T}}(B^{\pm}) > 22\,\GeV}$) is shown at the
bottom. The rapidity requirement is $|y(B^{\pm})|<2.3$.
The fits are used to extract
$N^{\text{reco}}(B^{+})+N^{\text{reco}}(B^{-})$ and its uncertainty
in each bin. The inset below each plot shows the bin-by-bin
difference between the data point and the value obtained from the
fit function divided by the quadrature sum of the statistical
uncertainty and the systematic uncertainties obtained from varying the fit model.
}
\label{fig:NrecoBpl2r}
\end{figure}
 
\begin{figure}[H]
\centering
\begin{tabular}{c}
\includegraphics[width=8.2 cm]{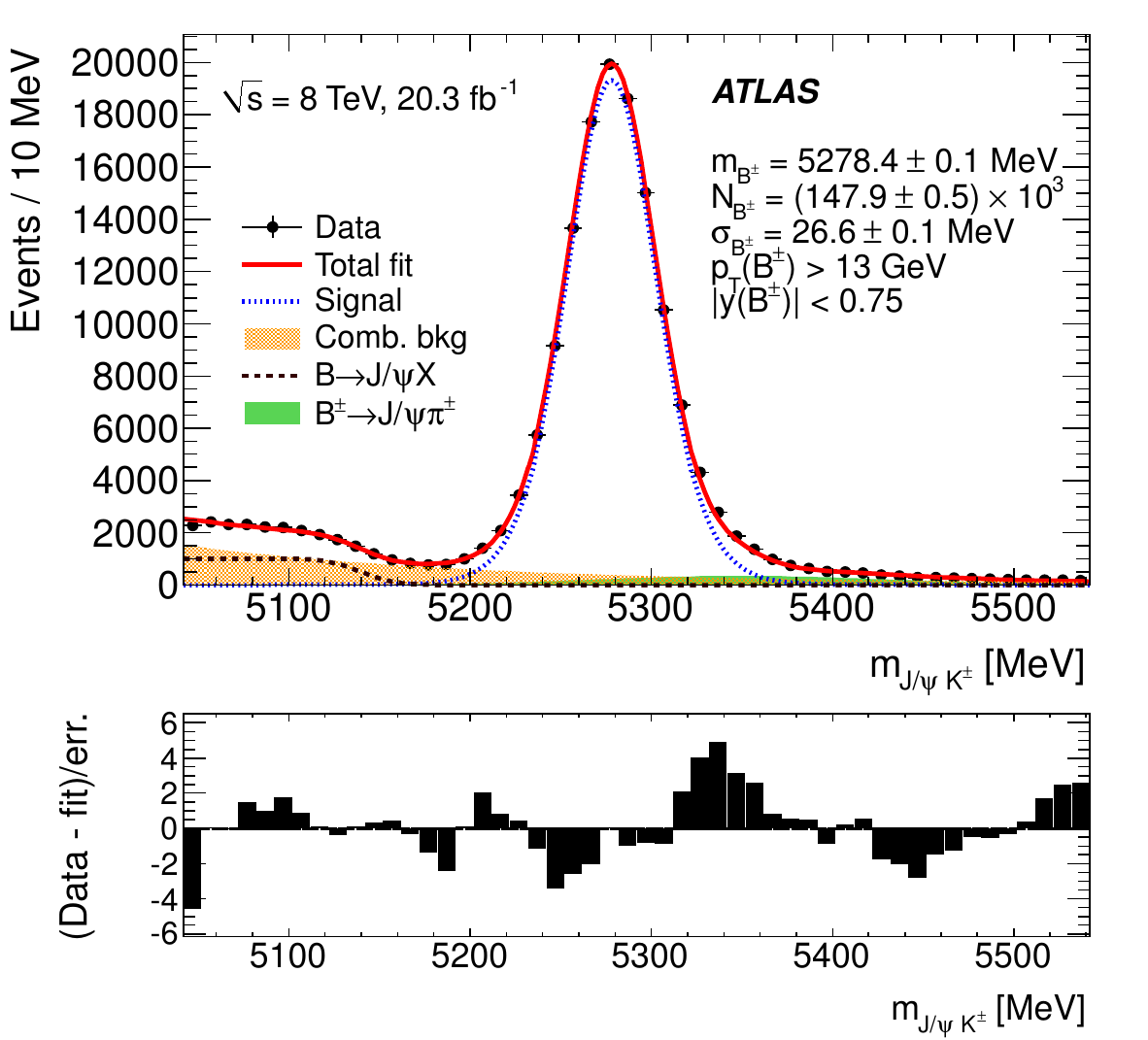} \\
\includegraphics[width=8.2 cm]{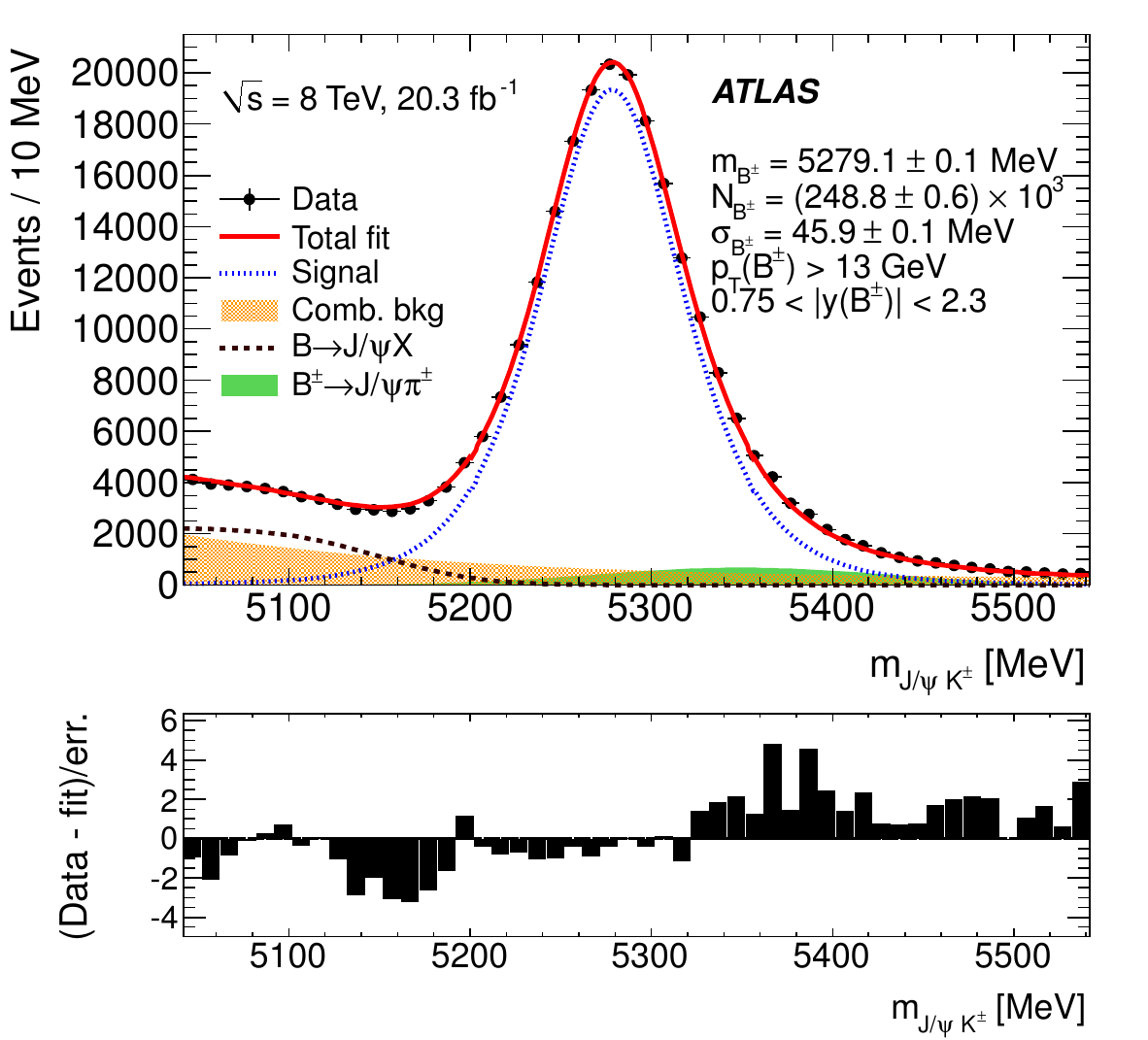} \\
\end{tabular}
\caption[The $B^{\pm}$ mass fit in two $|y(B^{\pm})|$ bins.]{The projection of the fit
of the invariant mass distribution for the $B^{\pm}$ meson.
``Signal'' stands for the signal component of the
fit, ``Comb.\ bkg'' stands for the combinatorial background
component of the fit, and ``Total fit'' stands for the fit to
the sum of the signal and all background components.
The fit
for the inner rapidity bin ($|y(B^{\pm})| < 0.75$) is shown at the
top, and that for the outer rapidity bin
($0.75 < |y(B^{\pm})| < 2.3$) is shown at the bottom.
The $p_{\mathrm{T}}$ requirement is
$p_{\mathrm{T}}(B^{\pm})>13$\,\GeV.
The fits are
used to extract
$N^{\text{reco}}(B^{+})+N^{\text{reco}}(B^{-})$ and its uncertainty
in each bin. The inset below each plot shows the bin-by-bin
difference between the data point and the value obtained from the
fit function divided by the quadrature sum of the statistical
uncertainty and the systematic uncertainties obtained from varying the fit model.
}
\label{fig:NrecoBpl2etar}
\end{figure}

\begin{figure}[H]
\centering
\begin{tabular}{c}
\includegraphics[width=8.2 cm]{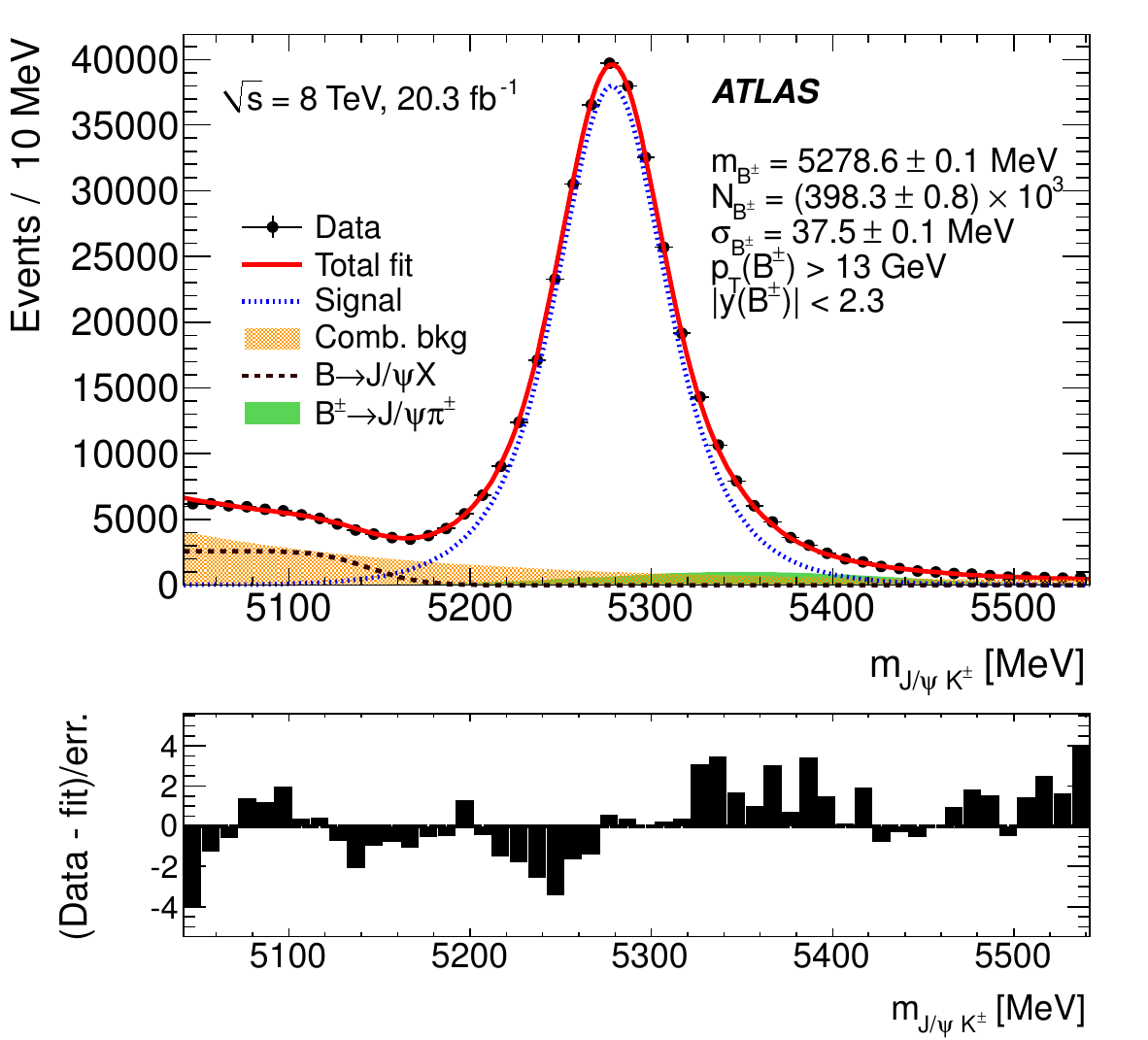}
\end{tabular}
\caption[The $B^{\pm}$ mass fit in two $p_{\mathrm{T}}(B^{\pm})$
bins.]{The projection of the fit of the invariant mass distribution for the $B^{\pm}$
meson using the complete dataset
($p_{\mathrm{T}}(B^{\pm}) > 13\,\GeV$, $|y(B^{\pm})|<2.3$).
``Signal'' stands for the signal component of the
fit, ``Comb.\ bkg'' stands for the combinatorial background
component of the fit, and ``Total fit'' stands for the fit to
the sum of the signal and all background components.
The fit is used to extract
$N^{\text{reco}}(B^{+})+N^{\text{reco}}(B^{-})$ and its uncertainty
in each bin. The inset below each plot shows the bin-by-bin
difference between the data point and the value obtained from the
fit function divided by the quadrature sum of the statistical
uncertainty and the systematic uncertainties obtained from varying the fit model.
}
\label{fig:NrecoBpl2etarFull}
\end{figure}

\begin{figure}[H]
\centering
\begin{tabular}{c}
\includegraphics[width=8.2 cm]{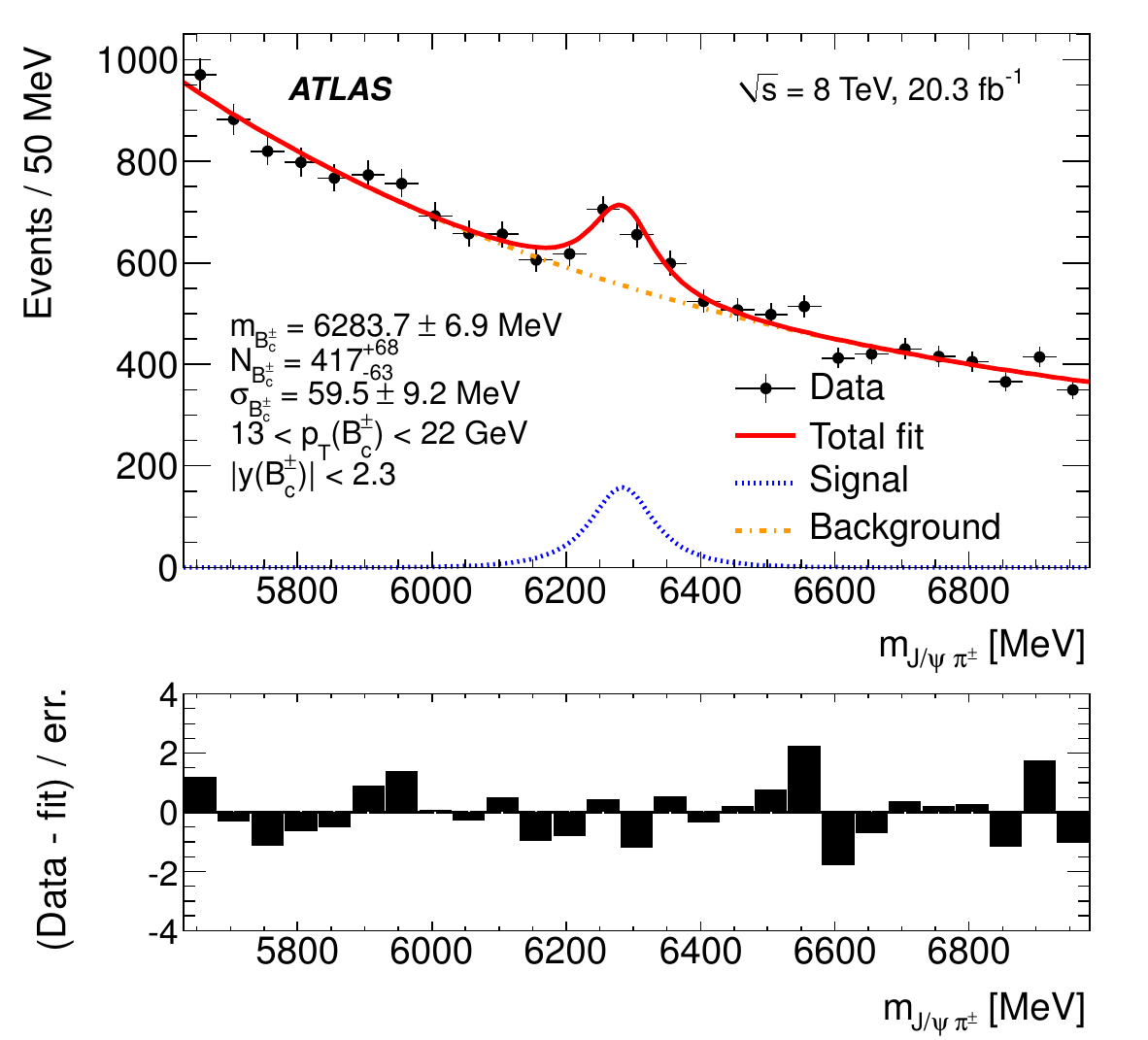}\\
\includegraphics[width=8.2 cm]{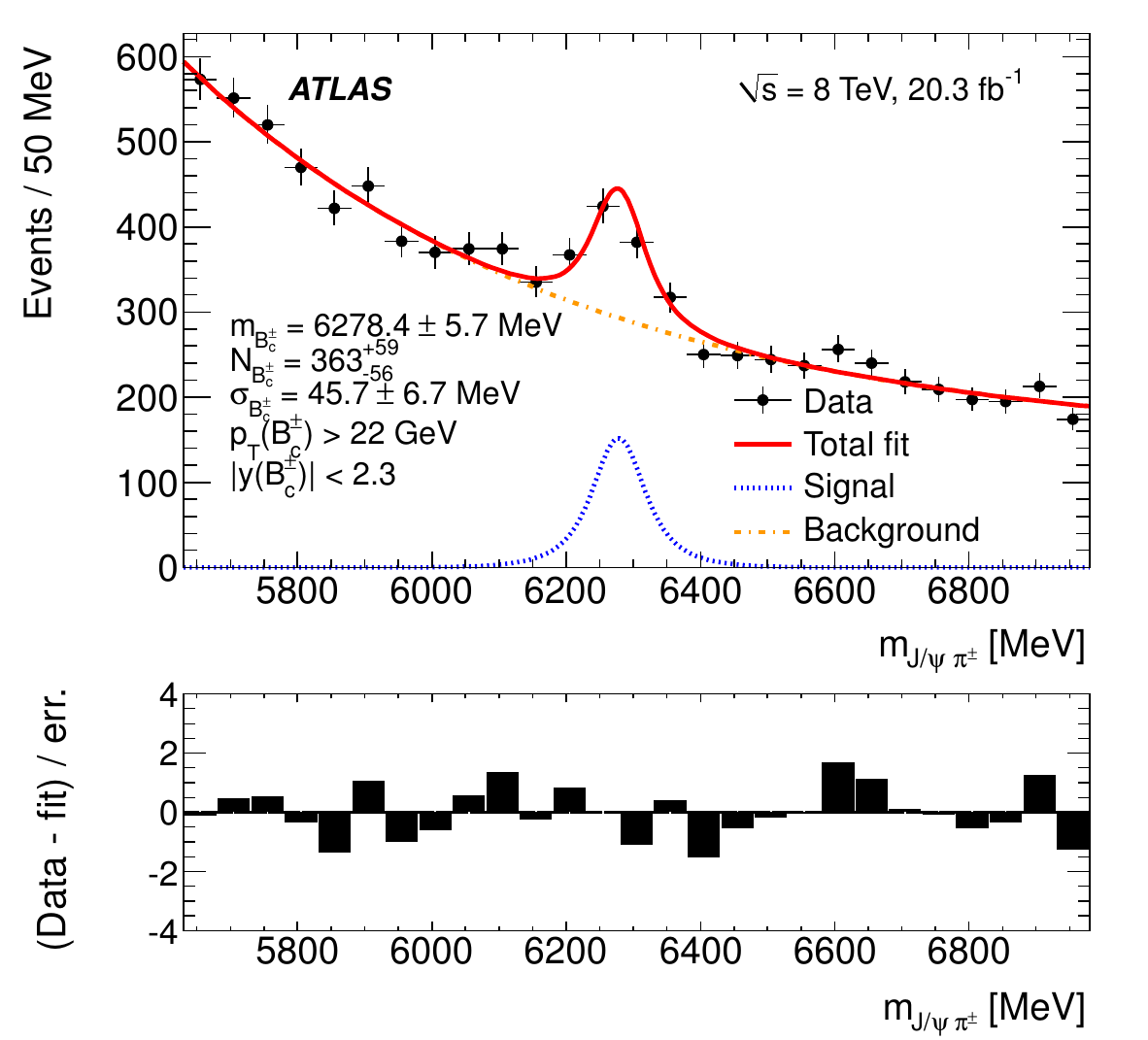}\\
\end{tabular}
\caption[The $B_{c}^{\pm}$ mass fit in two $p_{\mathrm{T}}(B^{\pm})$
bins.]{The projection of the fit of the invariant mass distribution for the $B_{c}^{\pm}$ meson.
``Signal'' stands for the signal component of the
fit, ``Background'' stands for the combinatorial background
component of the fit, and ``Total fit'' stands for the fit to
the sum of the signal and background components.
The fit for the low transverse momentum bin
($13\,\GeV < p_{\mathrm{T}}(B_{c}^{\pm}) < 22\,\GeV$)
is shown at the top, and that for the high transverse momentum bin
($p_{\mathrm{T}}(B_{c}^{\pm}) > 22\,\GeV$) at the
bottom. The rapidity requirement is $|y(B_{c}^{\pm})|<2.3$. The fits are used to extract
$N^{\text{reco}}(B_{c}^{+})+N^{\text{reco}}(B_{c}^{-})$ and its
uncertainty in each bin. The inset below each plot shows the bin-by-bin
difference between the data point and the value obtained from the
fit function divided by the quadrature sum of the statistical
uncertainty and the systematic uncertainties obtained from varying
the fit model.
}
\label{fig:NrecoBc12r}
\end{figure}

\begin{figure}[H]
\centering
\begin{tabular}{c}
\includegraphics[width=8.2 cm]{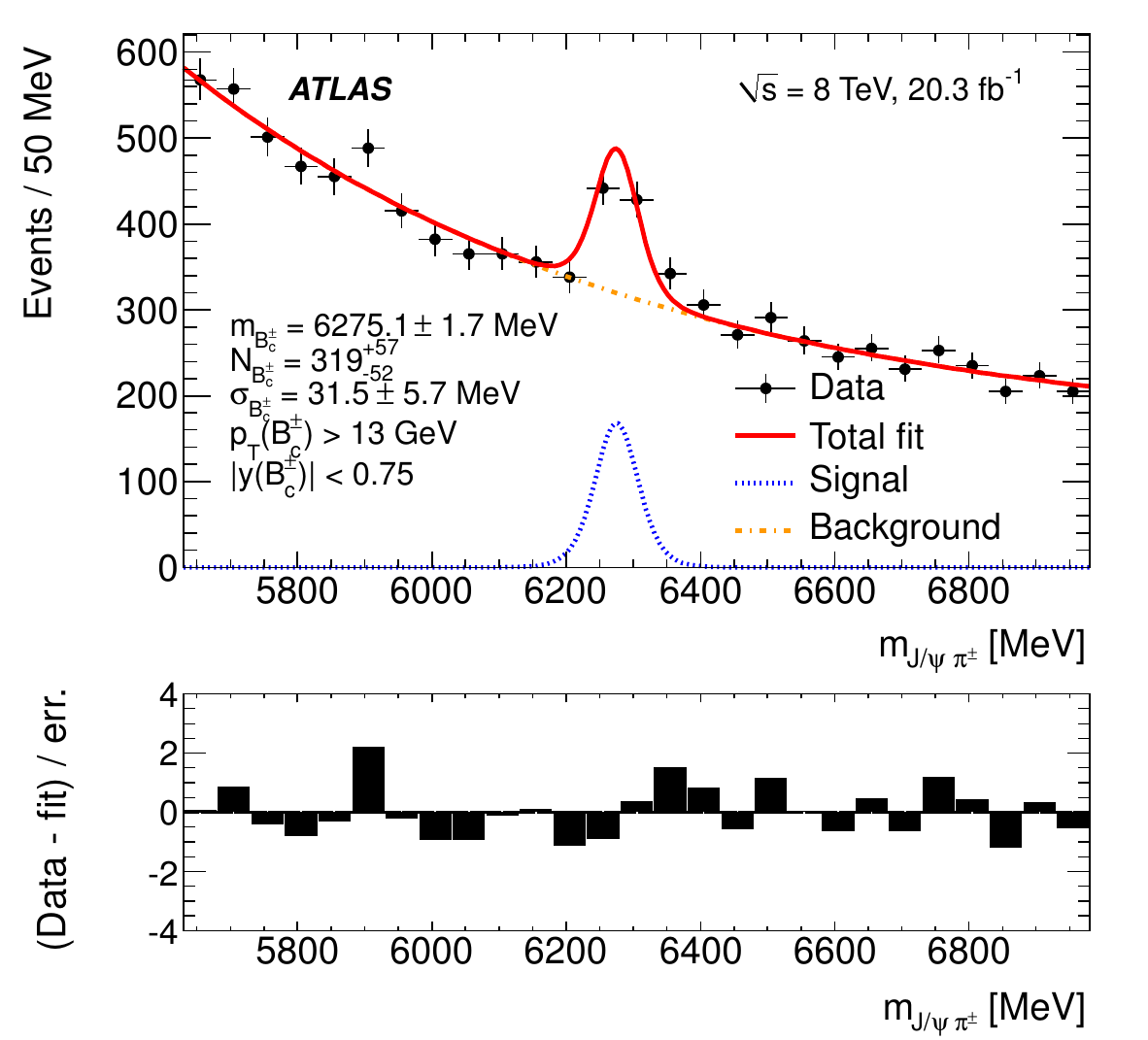}\\
\includegraphics[width=8.2 cm]{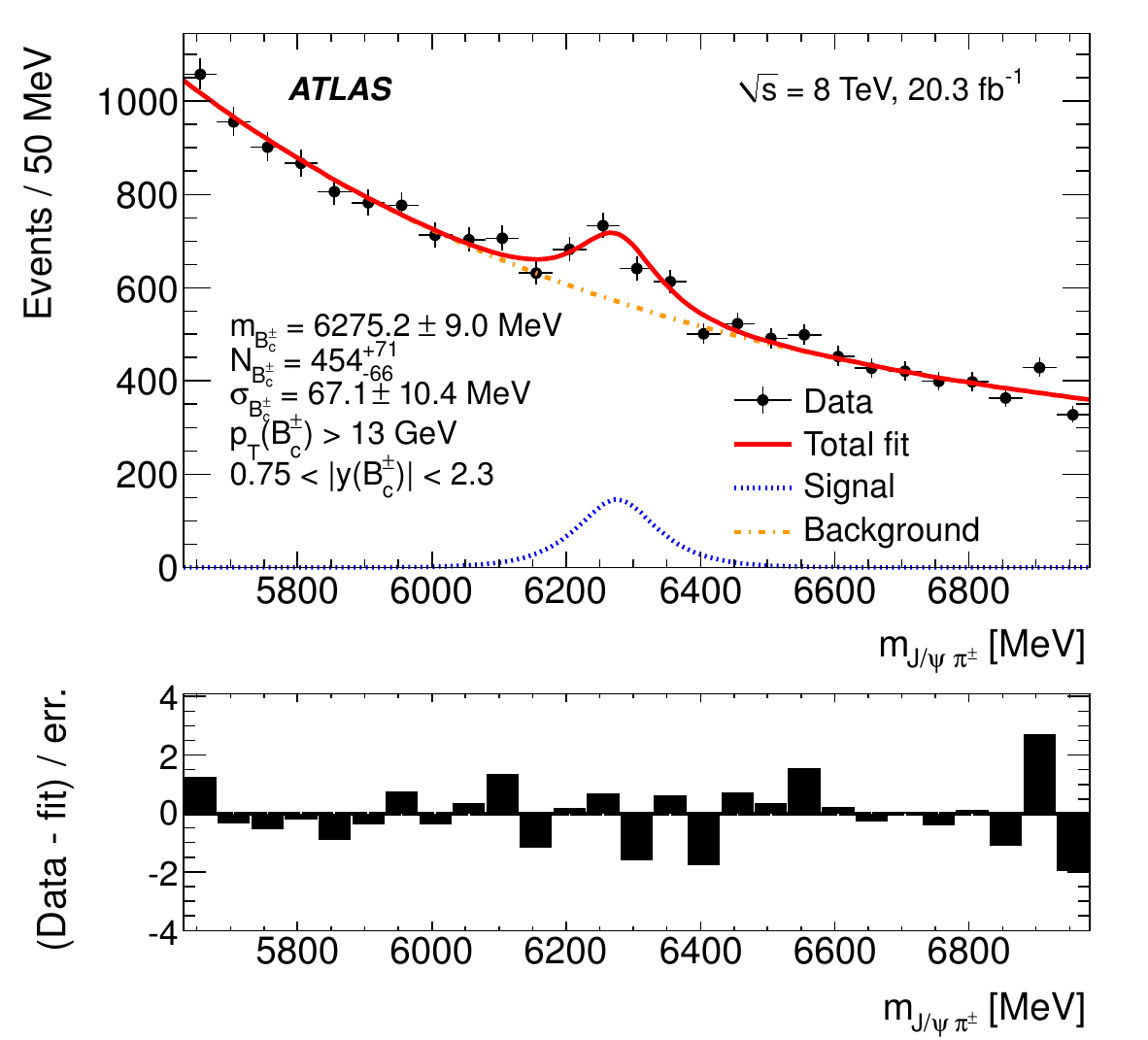}\\
\end{tabular}
\caption[The $B_{c}^{\pm}$ mass fit in two $p_{\mathrm{T}}(B_c^{\pm})$
bins.]{The projection of the fit of the invariant mass distribution for the $B_{c}^{\pm}$ meson.
``Signal'' stands for the signal component of the
fit, ``Background'' stands for the combinatorial background
component of the fit, and ``Total fit'' stands for the fit to
the sum of the signal and background components.
The
fit for the inner rapidity bin ($|y(B_{c}^{\pm})| < 0.75$) is shown
at the top, and that for the outer rapidity bin
($0.75 < |y(B_{c}^{\pm})| < 2.3$) at the bottom.
The $p_{\mathrm{T}}$ requirement is
$p_{\mathrm{T}}(B_{c}^{\pm})>13$\,\GeV.
The fits are used
to extract $N^{\text{reco}}(B_{c}^{+})+N^{\text{reco}}(B_{c}^{-})$ and
its uncertainty in each bin. The inset below each plot shows the bin-by-bin
difference between the data point and the value obtained from the
fit function divided by the quadrature sum of the statistical
uncertainty and the systematic uncertainties obtained from varying
the fit model.
}
\label{fig:NrecoBc12etar}
\end{figure}

\begin{figure}[H]
\centering
\begin{tabular}{c}
\includegraphics[width=8.2 cm]{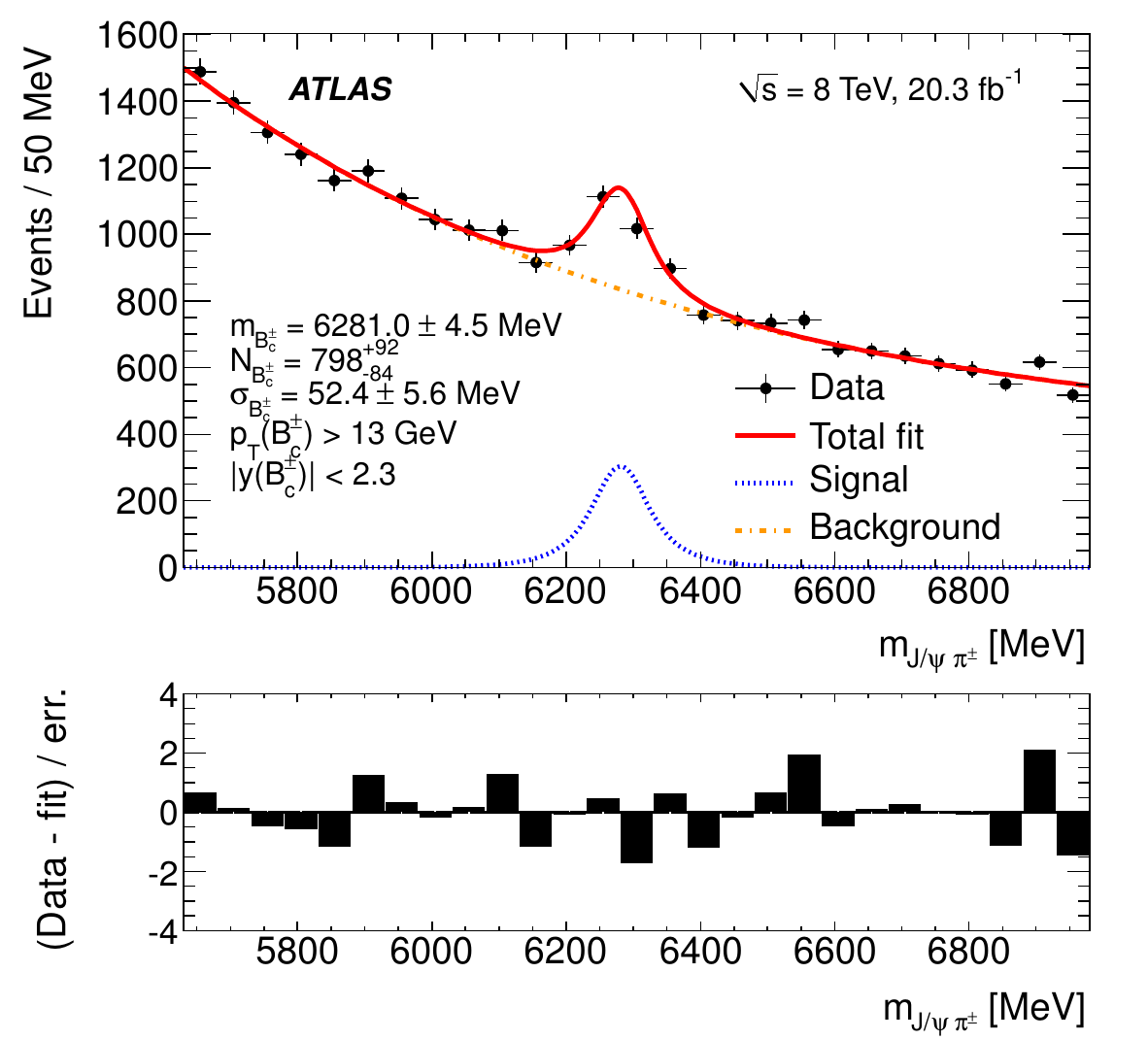}
\end{tabular}
\caption[The $B_{c}^{\pm}$ mass fit in two $p_{\mathrm{T}}(B^{\pm})$
bins.]{The projection of the fit of the invariant mass distribution for the $B_{c}^{\pm}$ meson using
the complete dataset
($p_{\mathrm{T}}(B_c^{\pm}) > 13\,\GeV$, $|y(B_c^{\pm})|<2.3$).
``Signal'' stands for the signal component of the
fit, ``Background'' stands for the combinatorial background
component of the fit, and ``Total fit'' stands for the fit to
the sum of the signal and background components.
The fit is used to extract
$N^{\text{reco}}(B_{c}^{+})+N^{\text{reco}}(B_{c}^{-})$ and its
uncertainty in each bin. The inset below each plot shows the bin-by-bin
difference between the data point and the value obtained from the
fit function divided by the quadrature sum of the statistical
uncertainty and the systematic uncertainties obtained from varying
the fit model.
}
\label{fig:NrecoBc12etarFull}
\end{figure}

The projections of the results of the invariant-mass fits of the
$B^{\pm}$ and the $B_c^{\pm}$ for the various $p_{\mathrm T}$ and $y$
bins considered in this measurement are given in
Figures~\ref{fig:NrecoBpl2r}--\ref{fig:NrecoBc12etarFull}. The inset below
each plot shows the bin-by-bin difference between each data point and
the value obtained from the fit function divided by the quadrature sum
of the statistical uncertainty and the systematic uncertainties obtained
from varying the fit model and discussed in Section~\ref{sec:systs}.
Tables~\ref{tab:fitsumBplus} and \ref{tab:fitsumBc} show a summary
of the main parameters of the fits.
 
\begin{table}[H]
\footnotesize
\begin{centering}
\caption{Summary of the main parameters of the $B^{\pm}$ fits. The uncertainties quoted are statistical.}
\label{tab:fitsumBplus}
\begin{tabular}{| l | c | c | c |}
\hline
Analysis bin & Fitted mass of the $B^{\pm}$ [\MeV] & Number of the $B^{\pm}$ candidates & $\sigma_m$ of the $B^{\pm}$ [\MeV]  \\
\hline
\hline
$p_{\mathrm T}(B) > 13$ \GeV, $|y(B)|<2.3$              & $5278.6 \pm 0.1$ & $(398.3\pm0.8)\times10^3$ & $37.5 \pm 0.1$ \\[1mm]
$13 < p_{\mathrm T}(B) < 22$ \GeV, $|y(B)|<2.3$      & $5278.5 \pm 0.1$ & $(207.6\pm0.6)\times10^3$ & $37.5 \pm 0.1$ \\[1mm]
$p_{\mathrm T}(B) > 22$ \GeV, $|y(B)|<2.3$               & $5278.8 \pm 0.1$ & $(190.9\pm0.6)\times10^3$ & $38.1 \pm 0.1$ \\[1mm]
$p_{\mathrm T}(B) > 13$ \GeV, $|y(B)| < 0.75$           & $5278.4 \pm 0.1$ & $(147.9\pm0.5)\times10^3$ & $26.6 \pm 0.1$ \\[1mm]
$p_{\mathrm T}(B) > 13$ \GeV, $0.75 < |y(B)| < 2.3$ & $5279.1 \pm 0.1$ & $(248.8\pm0.6)\times10^3$ & $45.9 \pm 0.1$ \\[1mm]
 
\hline
\end{tabular}
\end{centering}
\end{table}

\begin{table}[H]
\footnotesize
\begin{centering}
\caption{Summary of the main parameters of the $B_c^{\pm}$ fits. The uncertainties quoted are statistical.}
\label{tab:fitsumBc}
\begin{tabular}{| l | c | c | c |}
\hline
Analysis bin & Fitted mass of the $B_c^{\pm}$ [\MeV] & Number of the $B_c^{\pm}$ candidates & $\sigma_m$ of the $B_c^{\pm}$ [\MeV]  \\
\hline
\hline
$p_{\mathrm T}(B) > 13$ \GeV, $|y(B)|<2.3$              & $6281.0 \pm 4.5$ & $798^{+92}_{-84}$ & $52.4 \pm 5.6$ \\[1mm]
$13 < p_{\mathrm T}(B) < 22$ \GeV, $|y(B)|<2.3$      & $6283.7 \pm 6.9$ & $417^{+68}_{-63}$ & $59.5 \pm 9.2$ \\[1mm]
$p_{\mathrm T}(B) > 22$ \GeV, $|y(B)|<2.3$               & $6278.4 \pm 5.7 $ & $363^{+59}_{-56}$ & $45.7 \pm 6.7$ \\[1mm]
$p_{\mathrm T}(B) > 13$ \GeV, $|y(B)| < 0.75$           & $6275.1 \pm 1.7$ & $319^{+57}_{-52}$ & $31.5 \pm 5.7$ \\[1mm]
$p_{\mathrm T}(B) > 13$ \GeV, $0.75 < |y(B)| < 2.3$ & $6275.2 \pm 9.0$ & $454^{+71}_{-66}$ & $~~67.1 \pm 10.4$ \\[1mm]
 
\hline
\end{tabular}
\end{centering}
\end{table}

\section{Determination of the $B_c/B^{\pm}$ efficiency and acceptance ratio}
 
In order to correct the MC distributions to match the observed data,
a sPlot-based MC reweighting technique \cite{sPlot} is exploited.
This procedure is applied separately to the $B^{\pm}_c$ and $B^{\pm}$
candidates.
The following MC variables are reweighted: $p_{\mathrm T}$ of the $B$ candidate, $y$ of
the $B$ candidate, $\chi^2$ of the secondary vertex, and the
transverse impact parameter significance $d^0_{xy} /\sigma(d^0_{xy})$.
 
The reconstruction efficiencies are determined using the MC
samples for the $B_{c}^{\pm}$ and $B^{\pm}$ signals.
The efficiencies are averaged over each bin in $p_{\mathrm{T}}$
and each bin in $|y|$. They are calculated from the ratio of the
number of reconstructed MC events $N^{\text{reco}}_{\text{MC}}$ to
the number of generated MC events $N^{\text{gen}}_{\text{MC}}$ in
the associated bins in the fiducial region. Bin-to-bin migration is
included and found to be less than 0.1\%,
and the associated systematic uncertainties are significantly smaller
than uncertainties from other sources.
 
The efficiencies can be factorized into the product of the efficiency
of the $J/\psi$ trigger $\epsilon^{\text{trigger}}$, the efficiency of
the muon spectrometer $\epsilon^{\text{MS}}$, the efficiency of the
inner detector $\epsilon^{\text{ID}}$, the efficiency of fitting the
muon and hadron tracks to a common decay vertex
$\epsilon^{\text{vertex}}$, and the efficiency of the selection
criteria $\epsilon^{\text{selection}}$ \cite{Thessaloniki}:
\begin{equation*} \epsilon = \epsilon^{\text{trigger}}\cdot
\epsilon^{\text{MS}}(\mu^{+}) \cdot \epsilon^{\text{MS}}(\mu^{-})
\cdot \left(\epsilon^{\text{ID}}(\mu^{\pm})\right)^{2} \cdot
\epsilon^{\text{ID}}(X_h) \cdot \epsilon^{\text{vertex}}(B) \cdot
\epsilon^{\text{selection}}(B), \label{eqn:EffFact} \end{equation*}
where $\epsilon(X_h)$ is the $K^{\pm}$ efficiency in the $B^{\pm}$ channel and
$\pi^{\pm}$ efficiency in the $B_{c}^{\pm}$ channel, while $B$
stands for the $B^{\pm}$ and $B_c^{\pm}$ candidates.
Muon trigger efficiencies are calculated using the
method outlined in Ref.~\cite{psi2s}, which can be briefly summarized as
follows. The single-muon trigger efficiency is determined from
a tag-and-probe study of the $J/\psi$ and $\Upsilon$ dimuon decays in
Ref.~\cite{dimuon}. The
efficiency map is calculated as a function of $p_{\mathrm T}(\mu)$ and
$q \times \eta(\mu)$, where $q = \pm1$ is the electric charge of the
$\mu^{\pm}$, expressed in units of $e$.

Besides the product of two single-muon terms, the trigger efficiency
includes components that account for reductions in efficiency due to
closely spaced muons firing only a single RoI, for vertex quality and
opposite-sign charge requirements.

For the calculation of $\epsilon^{\text{MS}}(\mu^{\pm})$, the muon
reconstruction maps \cite{MCP} are used. These maps are based on a
sample of about two million $J/\psi \to \mu^+ \mu^-$ events collected
with unbiased triggers (single muonic and ``muon +
track''). The efficiency is measured in bins of $p_{\mathrm T}$ and $\eta$
using the same tag-and-probe method. Combined muons are used as the tag
muons and the ID tracks with the standard quality requirements are
used as probes.
The efficiency $\epsilon^{\text{ID}}(\mu^{\pm})$ of the muon track
reconstruction with the ID is conservatively taken to be $(99\pm1)$\%
\cite{XXX, YYY}.
 
The vertexing efficiency $\epsilon^{\text{vertex}}$ is estimated
with data and MC simulation for the $B^{\pm}$ and the $B^{\pm}_c$.
The efficiency ratio
$\epsilon^{\text{vertex}}(B^{\pm})/\epsilon^{\text{vertex}}(B^{\pm}_c)$
is found to be $1.01 \pm 0.01_{\text{stat}}$.

The MC samples ($B^{\pm}_c\to J/\psi \pi^{\pm}$ and
$B^{\pm}\to J/\psi K^{\pm}$) were generated with requirements on hadron
$p_{\mathrm T}$ of
$p_{\mathrm T} > 500$\,\MeV\ and $|\eta| < 2.5$. The requirements on muons
were $p_{\mathrm T} > 2.5$\,\GeV\ and $|\eta| < 2.7$. These are called the
minimal selection criteria (MSC).
To determine corrections to the efficiencies due to MSC,
a dedicated MC sample was produced for each channel
with no selection on pion, kaon, and muon momenta
and with a requirement that their absolute rapidity be less than 10.
The MC samples are corrected to take into
account the MSC, and these correction factors are
propagated to the analysis results. The following values are computed:
$f_{\text{cor}}(B_c^{\pm}) = N(B_c^{\pm})_{\text{MSC}}/N(B_c^{\pm})_{\text{no MSC}}$,
and
$f_{\text{cor}}(B^{\pm}) = N(B^{\pm})_{\text{MSC}}/N(B^{\pm})_{\text{no MSC}}$,
where $N(B)_{\text{no MSC}}$ and $N(B)_{\text{MSC}}$ stand for the numbers of
$B$ decays before and after applying the MSC, respectively,
in the $p_{\mathrm T}$ and $|y|$ bins used in the analysis.
These correction factors range from 8\% to
22\% depending on the $p_{\mathrm T}$ and $|y|$ of the $B$
candidates.
The value of the ratio $f_{\text{cor}}(B_c^{\pm})/f_{\text{cor}}(B^{\pm})$ is propagated
as a correction factor to the relative $B^{\pm}_c/B^{\pm}$ cross section.
The corrections obtained, along with their uncertainties,
are summarized in Table~\ref{tab:corr}. Each entry in the rightmost
column in Table~\ref{tab:corr} is then multiplied by the corresponding
value of the ratio of efficiencies
$\epsilon(B^{\pm})/\epsilon(B_{c}^{\pm})$.
The systematic uncertainty on the MSC procedure is estimated as the
difference between the raw MC prediction and the one reweighted using
the sPlot-based technique. The uncertainties in these
corrections and each of the uncertainties contributing to the
final efficiency ratios are added in quadrature.

\begin{table}[H]
\footnotesize
\begin{centering}
\caption{Summary of corrections due to the minimal selection criteria in
the MC simulation. The first uncertainty is statistical, the second
one is systematic.}
\label{tab:corr}
\begin{tabular}{| l | c | c | c |}
\hline
Analysis bin & Correction to the $B_c^{\pm}$ & Correction to the $B^{\pm}$ & Ratio of the corrections \\
\hline
\hline
$p_{\mathrm T}(B) > 13$ \GeV, $|y(B)|<2.3$                    & $0.0969 \pm 0.0004 \pm 0.0024$ & $0.0929 \pm 0.0004 \pm 0.0022$ & $0.959 \pm 0.006 \pm 0.024$   \\
$13 < p_{\mathrm T}(B) < 22$ \GeV, $|y(B)|<2.3$           & $0.0829 \pm 0.0004 \pm 0.0031$ & $0.0826 \pm 0.0004 \pm 0.0029 $ & $0.996 \pm 0.007 \pm 0.039$   \\
$p_{\mathrm T}(B) > 22$ \GeV, $|y(B)|<2.3$                    & $0.2164 \pm 0.0014 \pm 0.0018$ & $0.2213 \pm 0.0013 \pm 0.0017 $ & $1.023 \pm 0.009 \pm 0.009$   \\
$p_{\mathrm T}(B) > 13$ \GeV, $|y(B)| < 0.75$                & $0.0984 \pm 0.0007 \pm 0.0033$ & $0.0996 \pm 0.0007 \pm 0.0035 $ & $1.012 \pm 0.010 \pm 0.035$     \\
$p_{\mathrm T}(B) > 13$ \GeV, $0.75 < |y(B)| < 2.3$      & $0.0952 \pm 0.0005 \pm 0.0014$ & $0.0859 \pm 0.0005 \pm 0.0016 $  & $0.902 \pm 0.007 \pm 0.014$   \\
 
\hline
\end{tabular}
\end{centering}
\end{table}

The efficiency of the analysis selection criteria,
$\epsilon^{\text{selection}}$, derived from MC simulation, is incorporated into the final
efficiency ratios given below. The efficiency ratios
$\epsilon(B^{\pm})/\epsilon(B_{c}^{\pm})$, excluding the
MSC corrections, are found to be
$2.19\pm0.05$ for $13\,\GeV <
p_{\mathrm{T}}<22\,\GeV$,
$1.22\pm0.03$ for $p_{\mathrm{T}} >
22\,\GeV$,
$1.75\pm0.03$ for $p_{\mathrm{T}} >
13\,\GeV$,
$1.74\pm0.05$ for $|y|<0.75$, and
$1.76\pm0.04$ for
$0.75<|y|<2.3$ (see Section 7).
Here and below, when the range of a single variable is specified, it is
implicitly understood that the full range is selected for the other
variable, namely $p_{\mathrm T} > 13\,\GeV$ and $|y|<2.3$.
The reason for the efficiency ratios being larger than one is
primarily that the $B^{\pm}$ has a longer lifetime than the
$B^{\pm}_c$. Due to the shorter $B^{\pm}_c$ lifetime, the combination
of the $d^0_{xy}/\sigma (d^0_{xy})$ and $p_{\mathrm T}$(hadron)
selections criteria affects the $B_c^{\pm}$ more than the $B^{\pm}$,
this fact explains the factor two difference in the relative analysis
efficiency between the $p_{\mathrm T}$ bins.

\section{Systematic uncertainties}
\label{sec:systs}
A summary of all sources of uncertainty that contribute to the
analysis efficiency values is given in Table~\ref{tab:UncEff}.
These are absolute values, not percentages.
The absolute values for the efficiency ratios are presented in Section~6. They
propagate directly to the final results via Eq.~(\ref{eqn:crossSection}).
The uncertainty in the ratio of efficiencies
of detecting a kaon versus a pion is shown in Table~\ref{tab:UncEff}
in the row titled ``Tracking uncertainty''.
Tables~\ref{tab:UncSum}, \ref{tab:UncSum1}, and \ref{tab:UncSum2}
contain the systematic uncertainties related to the number of signal
events.
 
The systematic uncertainties of the efficiency ratios are primarily given
by the systematic uncertainties of $\epsilon^{\text{ID}}(X_h)$. They are
dominated by the material description in the simulation of the
detector \cite{EffUnc}.
The material density affects the $K^{\pm}$ and $\pi^{\pm}$ detection
in different ways.

The uncertainty in the efficiencies is due to the size of the
MC sample and systematic uncertainties in the event
counting. The precision of the efficiencies due to the size
of the MC sample is calculated according to Bernoulli
statistics. The uncertainty in the
probability of generated events to fall into a specific bin in
$p_{\mathrm{T}}$ or $|y|$ and the uncertainty in the probability of
reconstructing these events are added in quadrature.
 
The sPlot-based MC reweighting procedure produces an additional systematic
uncertainty that is estimated by varying the reweighted distributions
while preserving agreement with the data at the 1$\sigma$ level. The
maximum deviation for the analysis efficiency is taken as the systematic
uncertainty of the analysis efficiency derived from the sPlot-based MC
reweighting procedure. The uncertainties in the efficiency
ratios are found to be $0.03$ for $13\,\GeV <
p_{\mathrm{T}} < 22\,\GeV$, $0.03$ for $p_{\mathrm{T}} >
22\,\GeV$, $0.04$ for $p_{\mathrm{T}} > 13\,\GeV$,
$0.05$ for $|y|<0.75$, and $0.06$ for $0.75<|y|<2.3$.

\begin{table}[H]
\footnotesize
\begin{centering}
\caption[Summary of Uncertainties]{Summary of the absolute values of systematic
uncertainties for the analysis efficiency ratios.}
\label{tab:UncEff}
\resizebox{\columnwidth}{!}{
\begin{tabular}{|p{3.5cm}||c|c|c|c|c|}
\hline
Source of uncertainty & \multicolumn{5}{|c|} {Absolute value of the
uncertainty in the efficiency ratio} \\
\hline\hline
&  $p_{\mathrm{T}} > 13\,\GeV$
& $13 < p_{\mathrm{T}} < 22\,\GeV$ &
$p_{\mathrm{T}} > 22\,\GeV$ &
$ |y| < 0.75 $ &
$ 0.75 < |y| < 2.3 $ \\
\hline
Size of the MC samples and the event counting & 0.03 & 0.05 & 0.03 & 0.05 & 0.04 \\
\hline
sPlot-based MC reweighting procedure & 0.04 & 0.03 & 0.03 & 0.05 & 0.06 \\
\hline
Minimal selection criteria & 0.04 & 0.09 & 0.02 & 0.06 & 0.03 \\
\hline
Tracking uncertainty & 0.01 & 0.01 & 0.01 & 0.01 & 0.01 \\
\hline
\end{tabular}
} 
\end{centering}
\end{table}

The systematic uncertainties of the fitting procedure (which influence
the number of signal events) involve the choice of signal model and the
choice of background model. They are
estimated by fitting the invariant mass distributions of the $B_{c}^{\pm}$ and
$B^{\pm}$ using alternative models for the signal and background shapes
and varying the mass range of the fit. The sources of uncertainty are
treated as uncorrelated. The maximum deviations from
the nominal values when performing alternative fits for each
of the sources are added in quadrature to form the systematic
uncertainty of the fitted number of events.
 
The influence of the choice of signal model on the $B_{c}^{\pm}$
signal yield is estimated by replacing the Gaussian function by
a Crystal Ball function \cite{CrystalBall, CrystalBall2}
\begin{equation*}\mathcal{F}_{\mathrm{signal}}^{\text{Crystal
Ball}}(m_{J/\psi\pi^{\pm}}) \propto \begin{cases}
\exp\left[-(m_{J/\psi\pi^{\pm}}-m_{B_{c}^{\pm}})^{2}/(2\sigma_{\mathrm{CB}}^{2})\right],
& \mbox{for } m_{J/\psi
\pi^{\pm}}>m_{B_{c}^{\pm}}-\alpha\sigma_{\mathrm{CB}}\\
\frac{(n/\alpha)^{n}\exp(-\alpha^{2}/2)}{\left[(m_{B_{c}^{\pm}}-m_{J/\psi\pi^{\pm}})/\sigma_{\mathrm{CB}}+(n/\alpha)-\alpha\right]^{n}},
& \mbox{for } m_{J/\psi\pi^{\pm}} <
m_{B_{c}^{\pm}}-\alpha\sigma_{\mathrm{CB}} \end{cases}, \label{eqn:crystalBall}\end{equation*}
convolved with a Gaussian function. All the parameters in the
convolution are free and their values are obtained from the fit itself.
The resulting uncertainty in $N^{\text{reco}}(B_{c}^{+})+N^{\text{reco}}(B_{c}^{-})$
is $2.4\%$ for $13\,\GeV < p_{\mathrm{T}}(B_{c}^{\pm})<22\,\GeV$,
$1.1\%$ for $p_{\mathrm{T}}(B_{c}^{\pm}) > 22 \,\GeV$,
$2.5\%$ for $|y|<0.75$, $2.8\%$ for $0.75<|y|<2.3$,
and $2.4\%$ for the combined bin $p_{\mathrm{T}}(B_{c}^{\pm}) > 13\,\GeV$.
This comes solely from changing the signal model; the effect of changing the
background model is described next.
 
The uncertainty due to the choice of background model is estimated as
the maximum deviation from the nominal signal yield
$N^{\text{reco}}(B_{c}^{\pm})$ when the default model is replaced by
fourth-order Chebychev polynomials of the second kind.
The result is $1.7\%$ for $13\,\GeV <
p_{\mathrm{T}}(B_{c}^{\pm}) < 22\,\GeV$, $1.2\%$ for
$p_{\mathrm{T}}(B_{c}^{\pm}) > 22\,\GeV$, $2.8\%$ for
$|y|<0.75$, $1.3\%$ for $0.75<|y|<2.3$, and $2.9\%$ for the
inclusive bin $p_{\mathrm{T}}(B_{c}^{\pm}) > 13\,\GeV$.
 
To estimate the effect of neglecting the contributions of the Cabibbo-suppressed
(CS) decays $B_{c}^{\pm} \to J/\psi K^{\pm}$, the nominal fit model is
adapted to include a CS contribution and the difference with respect to the
nominal result is taken as a systematic uncertainty. In addition, to estimate
the effect of neglecting both the CS contribution and the partially
reconstructed $B^{\pm}_c$ decays (PRD), MC pseudo-experiments have been used.
They are generated starting from the default yield results of the
fits and adding CS and PRD components.
The CS component is estimated based on its branching fraction and
efficiency ratio to be $8.6\%$ of the signal yield~\cite{PDG_2018},
while a conservative estimation of the PRD component is obtained
from a fit to the data.
The generated samples are then fit both with the default configuration
(no CS and no PRD) and with a fit including the CS and PRD components
and the differences on the signal yields are recorded.

The uncertainty in the number of reconstructed $B^{\pm}$ events due to
the choice of signal model, estimated in the same way as for the
$B_{c}^{\pm}$, is found to be $0.1\%$ for $13\,\GeV <
p_{\mathrm{T}}(B^{\pm}) < 22\,\GeV$, $0.2\%$ for
$p_{\mathrm{T}}(B^{\pm}) > 22\,\GeV$, $0.1\%$ for
$|y|<0.75$, $0.2\%$ for $0.75<|y|<2.3$, and $0.1\%$ for the
combined bin $p_{\mathrm{T}}(B^{\pm}) > 13\,\GeV$. The
total systematic uncertainty of the $B^{\pm}$ fit is calculated by
adding the uncertainties associated with the mass range, the signal
model, and the background model in quadrature.
 
To determine the systematic uncertainty due to the choice of
background model for the $B^{\pm}$ fit, the exponential function
in Eq.~(\ref{eqn:expFunc1}) is replaced by fourth-order Chebychev polynomials
of the second kind.
The result is $0.2\%$ for $13\,\GeV <
p_{\mathrm{T}}(B^{\pm}) < 22\,\GeV$, $0.2\%$ for
$p_{\mathrm{T}}(B^{\pm}) > 22\,\GeV$, $0.2\%$ for
$|y|<0.75$, $0.2\%$ for $0.75<|y|<2.3$, and $0.1\%$ for the
inclusive bin $p_{\mathrm{T}}(B^{\pm}) > 13\,\GeV$. All
these contributions are summarized in Tables~\ref{tab:UncSum},
\ref{tab:UncSum1}, and \ref{tab:UncSum2} in the row labeled
``Background model of the fit.''
 
The CS contribution to the $B^{\pm}$ fit is estimated
by considering two extreme scenarios: no contribution and one that is twice
the nominal fraction obtained from the fit. The difference between these
scenarios is taken as the uncertainty in the number of candidates.
As a cross-check, the functional form of the $B^{\pm} \to J/\psi \pi^\pm$
component is varied to the sum of a Gaussian and an asymmetric Johnson
function~\cite{Johnson-1,Johnson} and the resulting effects are found to be
covered by the quoted systematic uncertainty.
 
Dimuon trigger efficiencies
are calculated for both decay modes, and relative trigger efficiencies
for both channels considered
are identical within the systematic uncertainty.  The residual minor
uncertainty is propagated to the final combined uncertainty of the
result.
 
The current world average uncertainty in the $B^\pm_{c}$ lifetime of
0.507~ps is 0.009~ps \cite{PDG_2018}, which corresponds to about 2\%,
while the uncertainty in the $B^{\pm}$ lifetime is four times smaller.
To analytically estimate the upper limit on the uncertainty due to the lifetime,
the $d^{0}_{xy}$ significance is studied and treated as 100\% correlated
with the lifetime. Using the limiting value for the selection
criterion, which is $d_{xy}^0/\sigma(d_{xy}^0) = 1.2$, the uncertainty
is obtained by multiplying 1.2 by 2\% to yield 0.024. Applying
$\pm$0.024 to the $d^{0}_{xy}$ significance, the resulting change
in the number of $B^\pm_{c}$ signal events is taken as the uncertainty
due to the lifetime. The absolute values of this uncertainty are shown in
Tables~\ref{tab:UncSum}--\ref{tab:UncSum2}. As a cross-check, the $B^{\pm}_c$
MC exclusive signal sample is reweighted in order to reflect the
$\pm2\%$ lifetime uncertainty mentioned above. The analysis
efficiencies are recalculated and the maximum deviations are found to
have smaller impact than those from the main method, so the largest deviations
obtained from the main method are used as an estimator of the uncertainty.
 
The central values of the integrated luminosity for the two $B$-meson datasets are
exactly the same, so the integrated luminosity cancels out completely in the ratio and
the luminosity uncertainty does not contribute to the
uncertainty of this measurement.
Removing combinations in which one of
the hadronic candidates is identified as a muon contributes an uncertainty
that is very small compared with the total systematic uncertainty and is
consequently neglected.
 
A summary of all sources of systematic uncertainties that  contribute
to the number of signal events is given in Tables~\ref{tab:UncSum}--\ref{tab:UncSum2}.
The different components of the
systematic uncertainty are added in quadrature.
The components of uncertainties correlated between the $B_c^{\pm}$ and
the $B^{\pm}$ are subtracted for the relative production cross section times
branching fractions presented in Section 8.
 
\begin{table}[H]
\footnotesize
\begin{centering}
\caption[Summary of Uncertainties]{Summary of all systematic uncertainties for the number of signal events in the two $p_{\mathrm T}$ bins.}
\label{tab:UncSum}
\resizebox{\columnwidth}{!}{
\begin{tabular}{|p{3.5cm}||c|c|c|c|}
\hline
Source of uncertainty & \multicolumn{4}{|c|} {Uncertainty value} \\
\hline\hline
& \multicolumn{2}{|c|}{$B_{c}^{\pm}$} & \multicolumn{2}{|c|}{$B^{\pm}$}\\
\hline
& $13\,\GeV < p_{\mathrm{T}} < 22\,\GeV$ &
$p_{\mathrm{T}} > 22\,\GeV$ &
$13\,\GeV < p_{\mathrm{T}} < 22\,\GeV$ &
$p_{\mathrm{T}} > 22\,\GeV$\\
\hline
Signal model of the fit & 2.4\% & 1.1\% & 0.1\% & 0.2\% \\
\hline
CS and PRD components & $^{+19.3\%}_{-2.4\%}$ & $^{+19.9\%}_{-2.4\%}$ & 0.5\% & 0.5\%  \\[3pt]
\hline
Background model of the fit & 1.7\% & 1.2\% & 0.2\% & 0.2\% \\
\hline
Trigger and reconstruction effects & 0.9\% & 0.8\% & 1.2\% & 1.2\% \\
\hline
$B$-meson lifetime uncertainty & 1.1\% & 0.9\% & $<0.1\%~~~~$ & $<0.1\%~~~~$ \\
\hline
 
\end{tabular}
} 
\end{centering}
\end{table}

\begin{table}[H]
\footnotesize
\centering
\caption[Summary of Uncertainties]{Summary of all systematic uncertainties for the number of signal events in the two $|y|$ bins.}
\label{tab:UncSum1}
\begin{tabular}{|p{3.5cm}||c|c|c|c|}
\hline
Source of uncertainty & \multicolumn{4}{|c|} {Uncertainty value} \\
\hline\hline
& \multicolumn{2}{|c|}{$B_{c}^{\pm}$} & \multicolumn{2}{|c|}{$B^{\pm}$}\\
\hline
& $ |y| < 0.75 $ &
$  0.75 < |y| < 2.3$ &
$ |y| < 0.75 $ &
$ 0.75 < |y| < 2.3 $\\
\hline
Signal model of the fit & 2.5\% & 2.8\% & 0.1\% & 0.2\% \\
\hline
CS and PRD components & $^{+11.2\%}_{-2.4\%}$ & $^{+23.2\%}_{-2.4\%}$ & 0.5\% & 0.5\% \\[3pt]
\hline
Background model of the fit & 2.8\% & 1.3\% & 0.2\% & 0.2\% \\
\hline
Trigger effects and reconstruction effects & 1.1\% & 1.0\% & 1.2\% & 1.1\% \\
\hline
$B$-meson lifetime uncertainty & 1.0\% & 0.9\%& $<0.1\%~~~~$  & $<0.1\%~~~~$  \\
\hline
\end{tabular}
\end{table}

\begin{table}[H]
\footnotesize
 
\centering
\caption[Summary of Uncertainties]{Summary of all systematic uncertainties for the number of signal events in the combined bin ($p_{\mathrm T} > 13$\,\GeV, $|y| < 2.3$).}
\begin{tabular}{|p{5.5cm}||c|c|}
\hline
Source of uncertainty & \multicolumn{2}{|c|} {Uncertainty value} \\
\hline\hline
& \multicolumn{1}{|c|}{$B_{c}^{\pm}$} & \multicolumn{1}{|c|}{$B^{\pm}$}\\
 
\hline
Signal model of the fit & 2.4\% & 0.1\%  \\
\hline
CS and PRD components & $^{+17.4\%}_{-2.4\%}$ & 0.5\% \\[3pt]
\hline
Background model of the fit & 2.9\% & 0.1\%  \\
\hline
Trigger effects and reconstruction effects & 0.9\% & 0.9\%  \\
\hline
$B$-meson lifetime uncertainty & 0.7\% & $<0.1\%~~~~$  \\
\hline
 
\end{tabular}
\label{tab:UncSum2}
\end{table}

\section{Results}
The yields $N^{\text{reco}}(B_{c}^{\pm})$ and
$N^{\text{reco}}(B^{\pm})$, and their statistical uncertainties are
extracted from the maximum-likelihood fits of the respective
invariant mass distributions. The differential relative production
cross sections times branching fractions are calculated according to
Eq.~(\ref{eqn:crossSection}) for all bins in $p_{\mathrm{T}}$ and
$|y|$.

The differential relative production cross section for the
inclusive selection containing all events in the range $p_{\mathrm{T}}
> 13\,\GeV$ and $|y| < 2.3$ is
\begin{equation*}
\left.\frac{\sigma(B_{c}^{\pm})\cdot \mathcal{B}(B_{c}^{\pm}\to J/\psi \pi^{\pm})}{\sigma(B^{\pm})\cdot \mathcal{B}(B^{\pm}\to J/\psi K^{\pm})}\right.
= (0.34\,\pm\,0.04\,_{\text{stat}}\,^{+0.06}_{-0.02}\,_{\text{syst}}\,\pm\,0.01_{\text{lifetime}})\%,\label{eqn:prodXsecAllPt}\end{equation*}
\begin{itemize}
\item for $13\,\GeV < p_{\mathrm{T}} < 22\,\GeV$, $|y| < 2.3$
is $(0.44\,\pm\,0.07\,_{\text{stat}}\,^{+0.09}_{-0.04}\,_{\text{syst}}\,\pm\,0.01_{\text{lifetime}})\%,$
\item for $p_{\mathrm{T}} > 22\,\GeV$,  $|y| < 2.3$
is $(0.24\,\pm\,0.04\,_{\text{stat}}\,^{+0.05}_{-0.01}\,_{\text{syst}}\,\pm\,0.01_{\text{lifetime}})\%,$
\item for $p_{\mathrm{T}} > 13\,\GeV$, $|y| < 0.75$
is $(0.38\,\pm\,0.06\,_{\text{stat}}\,^{+0.05}_{-0.04}\,_{\text{syst}}\,\pm\,0.01_{\text{lifetime}})\%,$
\item and for $p_{\mathrm{T}} > 13\,\GeV$, $0.75 < |y| < 2.3$
is $(0.29\,\pm\,0.05\,_{\text{stat}}\,^{+0.07}_{-0.02}\,_{\text{syst}}\,\pm\,0.01_{\text{lifetime}})\%.$
\end{itemize}

\begin{figure}[H]
\centering
\begin{tabular}{c c}
\includegraphics[width=8.0 cm]{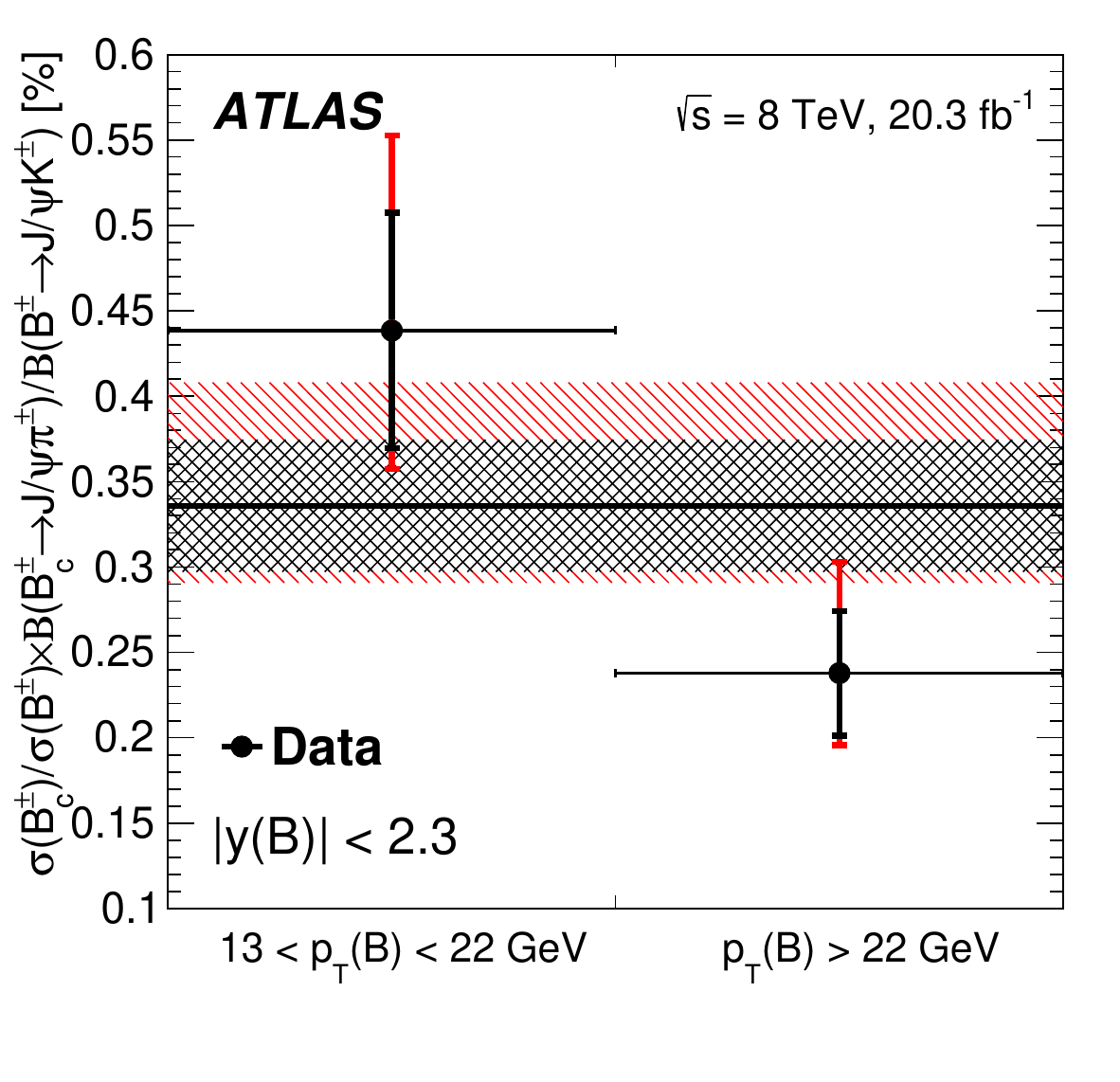} &
\hspace{-0.5cm}
\includegraphics[width=8.0 cm]{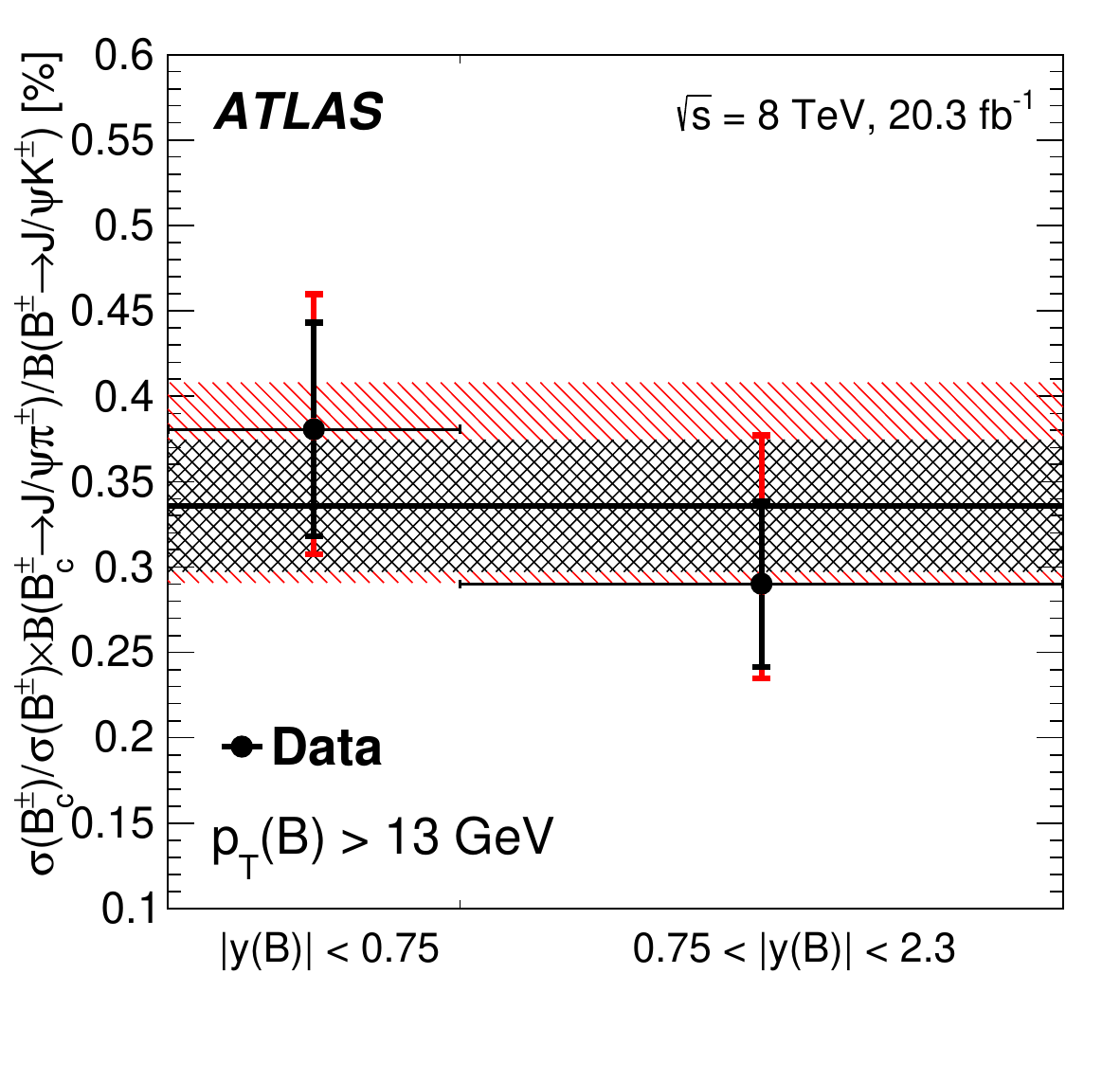}
\end{tabular}
\caption[]{Summary of the cross-section measurement presented in this
article. The left figure shows the production cross section
for the $B^{\pm}_c$ relative to the $B^{\pm}$ (times the corresponding
branching fractions) for two bins in $p_{\mathrm T}$
(black data points) and for the inclusive bin (horizontal band). The
right figure shows the measurement binned in rapidity (black data
points). On each point the horizontal bar indicates the bin
width. The vertical inner error bars on the data points indicate the
statistical uncertainty. The outer error bars indicate the size of the
quadrature sum of uncertainties from all sources: statistical,
systematic, and  lifetime. The double hatched error band indicates
statistical uncertainty, while the single hatched bands indicate the
quadrature sum of uncertainties from all sources. }
\label{fig:results}
\end{figure}

Figure~\ref{fig:results} summarizes the results of the measurement for
the $p_{\mathrm T}(B)$ and $|y(B)|$ bins as well as for the inclusive bin.
 
The differential measurement suggests a dependence on the transverse
momentum: the production cross section of the $B_{c}^{\pm}$ decreases
faster with $p_{\mathrm{T}}$ than the production cross section of the
$B^{\pm}$. No significant dependence on rapidity is observed.

\FloatBarrier

\section{Conclusion}
\label{sec:conclusion}
 
The production cross section of the $B_{c}^{\pm}$ meson relative to
the production cross section of the $B^{\pm}$ meson is measured
using $B_{c}^{\pm} \to J/\psi \pi^{\pm}$ and $B^{\pm} \to J/\psi
K^{\pm}$ decays reconstructed by the ATLAS detector analyzing $pp$
collisions at $\sqrt{s}$ = 8\,\TeV\ delivered by the LHC in 2012.
The data used for this study correspond to an integrated luminosity of
$20.3\,\mathrm{fb}^{-1}$.
The relative cross section times branching fraction for
the full range $p_{\mathrm T} > 13$\,\GeV\ and $|y| < 2.3$ is
$(0.34\,\pm\,0.04_{\text{stat}}\,^{+0.06}_{-0.02}\,_{\text{sys}}\,\pm\,0.01_{\text{lifetime}})\%$.
The ratio of the $B_c$ to $B^{\pm}$ cross sections is measured
in two intervals of transverse momentum and rapidity of the $B$-meson
candidates.
The differential measurement suggests a dependence on the transverse
momentum: the production cross section of the $B_{c}^{\pm}$ meson decreases
faster with $p_{\mathrm{T}}$ than the production cross section of the
$B^{\pm}$ meson. No significant dependence on rapidity is observed.
 
\clearpage
\section*{Acknowledgments}
This paper is dedicated to the memory of Konstantin Toms who has been
the main force behind this analysis and did not live to see it
published.

We thank CERN for the very successful operation of the LHC, as well as the
support staff from our institutions without whom ATLAS could not be
operated efficiently.
 
We acknowledge the support of ANPCyT, Argentina; YerPhI, Armenia; ARC, Australia; BMWFW and FWF, Austria; ANAS, Azerbaijan; SSTC, Belarus; CNPq and FAPESP, Brazil; NSERC, NRC and CFI, Canada; CERN; ANID, Chile; CAS, MOST and NSFC, China; COLCIENCIAS, Colombia; MSMT CR, MPO CR and VSC CR, Czech Republic; DNRF and DNSRC, Denmark; IN2P3-CNRS and CEA-DRF/IRFU, France; SRNSFG, Georgia; BMBF, HGF and MPG, Germany; GSRT, Greece; RGC and Hong Kong SAR, China; ISF and Benoziyo Center, Israel; INFN, Italy; MEXT and JSPS, Japan; CNRST, Morocco; NWO, Netherlands; RCN, Norway; MNiSW and NCN, Poland; FCT, Portugal; MNE/IFA, Romania; JINR; MES of Russia and NRC KI, Russian Federation; MESTD, Serbia; MSSR, Slovakia; ARRS and MIZ\v{S}, Slovenia; DST/NRF, South Africa; MICINN, Spain; SRC and Wallenberg Foundation, Sweden; SERI, SNSF and Cantons of Bern and Geneva, Switzerland; MOST, Taiwan; TAEK, Turkey; STFC, United Kingdom; DOE and NSF, United States of America. In addition, individual groups and members have received support from BCKDF, CANARIE, Compute Canada, CRC and IVADO, Canada; Beijing Municipal Science \& Technology Commission, China; COST, ERC, ERDF, Horizon 2020 and Marie Sk{\l}odowska-Curie Actions, European Union; Investissements d'Avenir Labex, Investissements d'Avenir Idex and ANR, France; DFG and AvH Foundation, Germany; Herakleitos, Thales and Aristeia programmes co-financed by EU-ESF and the Greek NSRF, Greece; BSF-NSF and GIF, Israel; La Caixa Banking Foundation, CERCA Programme Generalitat de Catalunya and PROMETEO and GenT Programmes Generalitat Valenciana, Spain; G\"{o}ran Gustafssons Stiftelse, Sweden; The Royal Society and Leverhulme Trust, United Kingdom.
 
The crucial computing support from all WLCG partners is acknowledged gratefully, in particular from CERN, the ATLAS Tier-1 facilities at TRIUMF (Canada), NDGF (Denmark, Norway, Sweden), CC-IN2P3 (France), KIT/GridKA (Germany), INFN-CNAF (Italy), NL-T1 (Netherlands), PIC (Spain), ASGC (Taiwan), RAL (UK) and BNL (USA), the Tier-2 facilities worldwide and large non-WLCG resource providers. Major contributors of computing resources are listed in Ref.~\cite{ATL-SOFT-PUB-2020-001}.
 

\printbibliography

\clearpage 
 
\begin{flushleft}
\hypersetup{urlcolor=black}
{\Large The ATLAS Collaboration}

\bigskip

\AtlasOrcid{M.~Aaboud}$^\textrm{\scriptsize 35e}$,    
\AtlasOrcid[0000-0002-6665-4934]{G.~Aad}$^\textrm{\scriptsize 102}$,    
\AtlasOrcid[0000-0002-5888-2734]{B.~Abbott}$^\textrm{\scriptsize 129}$,    
\AtlasOrcid[0000-0002-7248-3203]{D.C.~Abbott}$^\textrm{\scriptsize 103}$,    
\AtlasOrcid{O.~Abdinov}$^\textrm{\scriptsize 13,*}$,    
\AtlasOrcid[0000-0002-2788-3822]{A.~Abed~Abud}$^\textrm{\scriptsize 71a,71b}$,    
\AtlasOrcid[0000-0002-2987-4006]{D.K.~Abhayasinghe}$^\textrm{\scriptsize 94}$,    
\AtlasOrcid[0000-0002-8496-9294]{S.H.~Abidi}$^\textrm{\scriptsize 168}$,    
\AtlasOrcid[0000-0002-8279-9324]{O.S.~AbouZeid}$^\textrm{\scriptsize 40}$,    
\AtlasOrcid{N.L.~Abraham}$^\textrm{\scriptsize 157}$
% ...
\bigskip
\\

% ...
$^{13}$Institute of Physics, Azerbaijan Academy of Sciences, Baku; Azerbaijan.\\
% ...
$^{35}$$^{(a)}$Facult\'e des Sciences Ain Chock, R\'eseau Universitaire de Physique des Hautes Energies - Universit\'e Hassan II, Casablanca;$^{(b)}$Facult\'{e} des Sciences, Universit\'{e} Ibn-Tofail, K\'{e}nitra;$^{(c)}$Facult\'e des Sciences Semlalia, Universit\'e Cadi Ayyad, LPHEA-Marrakech;$^{(d)}$Moroccan Foundation for Advanced Science Innovation and Research (MAScIR), Rabat;$^{(e)}$LPMR, Facult\'e des Sciences, Universit\'e Mohamed Premier, Oujda;$^{(f)}$Facult\'e des sciences, Universit\'e Mohammed V, Rabat; Morocco.\\
% ...
$^{40}$Niels Bohr Institute, University of Copenhagen, Copenhagen; Denmark.\\
% ...
$^{71}$$^{(a)}$INFN Sezione di Pavia;$^{(b)}$Dipartimento di Fisica, Universit\`a di Pavia, Pavia; Italy.\\
% ...
$^{94}$Department of Physics, Royal Holloway University of London, Egham; United Kingdom.\\
% ...
$^{102}$CPPM, Aix-Marseille Universit\'e, CNRS/IN2P3, Marseille; France.\\
$^{103}$Department of Physics, University of Massachusetts, Amherst MA; United States of America.\\
% ...
$^{129}$Homer L. Dodge Department of Physics and Astronomy, University of Oklahoma, Norman OK; United States of America.\\
% ...
$^{157}$Department of Physics and Astronomy, University of Sussex, Brighton; United Kingdom.\\
% ...
$^{168}$Department of Physics, University of Toronto, Toronto ON; Canada.\\
% ...
$^{a}$ Also at Borough of Manhattan Community College, City University of New York, New York NY; United States of America.\\
$^{b}$ Also at Centre for High Performance Computing, CSIR Campus, Rosebank, Cape Town; South Africa.\\
$^{c}$ Also at CERN, Geneva; Switzerland.\\
$^{d}$ Also at CPPM, Aix-Marseille Universit\'e, CNRS/IN2P3, Marseille; France.\\
$^{e}$ Also at D\'epartement de Physique Nucl\'eaire et Corpusculaire, Universit\'e de Gen\`eve, Gen\`eve; Switzerland.\\
$^{f}$ Also at Departament de Fisica de la Universitat Autonoma de Barcelona, Barcelona; Spain.\\
% ...
$^{*}$ Deceased

\end{flushleft}


\end{document}